\documentclass{article}
\pdfoutput=1

\usepackage{jheppub}
\notoc

\usepackage{amssymb}
\usepackage{tikz}
\usepackage{amsmath}
\usepackage{hyperref}

\usepackage[numbers,sort&compress]{natbib} 

\newcommand{\ka}{\kappa}

\newcommand{\de}{\delta} 

\newcommand{\ba}{\begin{align}}
\newcommand{\ea}{\end{align}}	
\newcommand{\eref}[1]{Eq.~(\ref{#1})}
\newcommand{\fref}[1]{Fig.~\ref{#1}}
\newcommand{\nnnl}{\nonumber\\}	

\newcommand{\beq}{\begin{eqnarray}}
\newcommand{\eeq}{\end{eqnarray}}

\title{Two- and three-point functions in two-dimensional Landau-gauge Yang-Mills theory: Continuum results}

\author[a]{Markus Q. Huber,}

\author[b]{Axel Maas,}

\author[a]{Lorenz von Smekal}

\affiliation[a]{Institut f\"ur Kernphysik, Technische Universit\"at Darmstadt, Schlossgartenstr. 2, 64289 Darmstadt, Germany}

\affiliation[b]{Institute for Theoretical Physics, Friedrich-Schiller-University Jena, Max-Wien-Platz 1, D-07743 Jena, Germany}

\emailAdd{markus.huber@physik.tu-darmstadt.de}
\emailAdd{axel.maas@uni-jena.de}
\emailAdd{lorenz.smekal@physik.tu-darmstadt.de}

\abstract{
We investigate the Dyson-Schwinger equations for the gluon and ghost
propagators and the ghost-gluon vertex of Landau-gauge gluodynamics in
two dimensions. While this simplifies some aspects of the calculations as
compared to three and four dimensions, new complications arise due to
a mixing of different momentum regimes. As a result, the solutions for
the propagators are more sensitive to changes in the three-point
functions and the ans\"atze used for them at the leading order in a
vertex expansion. Here, we therefore go beyond this common truncation
by including the ghost-gluon vertex self-consistently for the first
time, while using a model for the three-gluon vertex which reproduces
the known infrared asymptotics and the zeros at intermediate momenta
as observed on the lattice. A separate computation of the three-gluon
vertex from the results is used to confirm the stability of this
behavior a posteriori. We also present further arguments for the absence of the
decoupling solution in two dimensions. Finally, we show how in general 
the infrared exponent $\ka$ of the scaling solutions in two, three and
four dimensions can be changed by allowing an angle dependence and thus
an essential singularity of the ghost-gluon vertex in the infrared.} 

\keywords{Green functions, Yang-Mills theory, two dimensional quantum field theory, infrared behavior}

\begin{document}
\maketitle

\section{Introduction}

Quantum chromodynamics (QCD) is the theory of quarks and gluons. It is well understood in the perturbative regime, but phenomena like confinement and chiral symmetry breaking are intrinsically non-perturbative and their investigation therefore requires adequate methods. Some of these effects are believed to be present already for the gluonic sector alone, i.~e., Yang-Mills theory, on which we will focus here. One approach to understand the non-perturbative properties is based on the correlation functions, which are also used as input in many phenomenological applications. However, correlation functions are in general gauge-dependent.

A common gauge choice is the Landau gauge, for which propagators and vertices have been calculated with several methods, see, for instance, \cite{vonSmekal:1997vx,vonSmekal:1997is,Zwanziger:2001kw,Lerche:2002ep,Zwanziger:2002ia,Fischer:2002hn,Pawlowski:2003hq,Zwanziger:2003cf,Alkofer:2004it,Silva:2004bv,Bogolubsky:2007ud,Oliveira:2007dy,Cucchieri:2007md,Dudal:2007cw,Dudal:2008sp,Boucaud:2008ji,Aguilar:2008xm,Cucchieri:2008qm,Alkofer:2008dt,Alkofer:2008jy,Fischer:2008uz,vonSmekal:2008ws,Fischer:2009tn,Huber:2009tx,Zwanziger:2010iz}, summarized in the recent reviews \cite{Binosi:2009qm,Boucaud:2011ug,Maas:2011se,Vandersickel:2012tz}. For various reasons investigations were extended from four also to three \cite{Zwanziger:2001kw,Lerche:2002ep,Maas:2004se,Huber:2007kc,Dudal:2008rm,Cucchieri:2008qm,Aguilar:2010zx} and two dimensions \cite{Zwanziger:2001kw,Lerche:2002ep,Huber:2007kc,Maas:2007uv,Dudal:2008xd,Cucchieri:2011um,Cucchieri:2012cb}. One of the main motivations for this is that lattice calculations become much cheaper in lower dimensions so one can more easily reach low momenta. An approach complementary to lattice calculations is based on functional equations \cite{Pawlowski:2010ht} like functional renormalization group equations, see, e.~g., \cite{Berges:2000ew,Pawlowski:2005xe,Gies:2006wv,Rosten:2010vm}, or Dyson-Schwinger equations (DSEs), see, e.~g., \cite{Roberts:1994dr,Roberts:2000aa,Alkofer:2000wg,Fischer:2006ub,Alkofer:2008nt,Binosi:2009qm,Maas:2011se}, for which large scale separations are less problematic. However, since they consist of an infinite tower of equations, truncations are required.

The situation in four dimensions is interesting, because with continuum methods one finds two types of solutions, while lattice methods yield only one. The two types are called decoupling and scaling solutions. The former \cite{Cucchieri:2007md,Bogolubsky:2007ud,Aguilar:2008xm,Boucaud:2008ji,Fischer:2008uz, Alkofer:2008jy} is characterized by a gluon propagator that becomes finite at zero momentum while the ghost propagator behaves like a massless particle. On the other hand, the scaling solution \cite{vonSmekal:1997vx,vonSmekal:1997is,Zwanziger:2001kw,Lerche:2002ep,Fischer:2008uz, Alkofer:2008jy} features an infrared (IR) vanishing gluon propagator and an IR enhanced ghost propagator. On the lattice only the decoupling type of solution is seen in four and three dimensions. Currently there does not exist a consensus in the community if only one solution is physical or both types are valid and possibly correspond to different non-perturbative completions of the Landau gauge, see \cite{Maas:2011se} for a compilation of the current state of affairs. The situation in three dimensions is essentially the same.

On the other hand, the two dimensional case is different: Lattice calculations \cite{Maas:2007uv,Cucchieri:2011um} seem not to find the decoupling type solution, but the results resemble more the scaling type solution. However, studies in the strong coupling limit, $\beta\rightarrow 0$, show that the situation is not completely clear, since the analytically known scaling relation for the IR exponents is not fulfilled for $\beta=0$ and the renormalized Landau gauge coupling does not approach a momentum-independent value \cite{Maas:2009ph}. The results for the dressing functions at finite $\beta$ fit results from analytic analyses using functional methods reasonably well \cite{Zwanziger:2001kw,Lerche:2002ep,Huber:2007kc,Cucchieri:2012cb}, but an agreement between the IR exponents extracted from the lattice data \cite{Maas:2007uv,Maas:2009ph,Cucchieri:2011um,Cucchieri:2011ig} is not reached. Results obtained within the Gribov-Zwanziger framework \cite{Dudal:2008xd} corroborate the non-existence of the decoupling type solution as does Ref. \cite{Cucchieri:2012cb}. In the latter also a bound on the gluon propagator at zero momentum incompatible with this type of solution was derived. Recently, an IR bound was derived from the restriction to the Gribov region in ref. \cite{Zwanziger:2012xg}.

In the present work we extend the currently available analytic results from DSEs in two dimensions \cite{Zwanziger:2001kw,Lerche:2002ep,Huber:2007kc,Cucchieri:2012cb} and present the first solutions at all momenta. We confirm the absence of the decoupling solution from Dyson-Schwinger equations in agreement with lattice and prior results also when taking the full momentum range into account. Although the underlying equations seem easier to solve in two dimensions than in four dimensions, for example, because no renormalization has to be performed, we find that they have their own intricacies. The reason is that in two dimensions different momentum regions influence each other. Especially the mid-momentum behavior, which is where the truncations of DSEs have the biggest effects, is very important in order to obtain correct results. This can be uniquely traced back to the subtle cancellations required for the scaling solution in two dimensions. As a consequence, it is non-trivial in two dimensions to find a viable truncation which works over the whole momentum range from the deep infrared to the ultraviolet. We will discuss several model ans\"atze for the vertices to deal with this issue. As a final step we also include the ghost-gluon vertex to the set of equations to be solved numerically. In this system of equations the only undetermined quantity is the three-gluon vertex, which we adjust so as to obtain the correct ghost UV behavior. 

We will explain the setup of the equations and fix our notation in Sec.~\ref{sec:YM2d}. Then we will consider the ghost DSE numerically in Sec.~\ref{sec:gh-DSE}. In Sec.~\ref{sec:IR-anal} we present analytic results, and the coupled system of the propagators DSEs is investigated numerically in Sec.~\ref{sec:gh+gl}. In Sec.~\ref{sec:ghgvert} we present the results of the enlarged system of propagators and ghost-gluon vertex. We conclude with a summary in Sec.~\ref{sec:conclusions}. Two appendices contain details on the integral kernels and results for the three-gluon vertex.

\section{Dyson-Schwinger equations of two-dimensional Yang-Mills theory}
\label{sec:YM2d}

The Euclidean Langrangian of Yang-Mills theory fixed to the Landau gauge reads
\begin{align}
\mathcal{L}&=\frac1{4}F_{\mu\nu}^aF_{\mu\nu}^a+\frac{1}{2\xi}(\partial A^a)^2-\bar{c}^a\,M^{ab}\, c^b,
\end{align}
where $F_{\mu\nu}^a$ is a component of the field strength tensor $F_{\mu\nu}=F_{\mu\nu}^a T^a$ with the Hermitian generators $T^a$ of the gauge group and $M^{ab}$ is the Faddeev-Popov operator:
\begin{align}
F_{\mu\nu}^a&:=\partial_\mu A_\nu^a -\partial_\nu A_\mu^a +i \,g\,f^{abc}A_\mu^b A_\nu^c,\\
 M^{ab}&:=-\de^{ab}\partial^2-g\,f^{abc}\partial_\mu A_\mu^c.
\end{align}
The DSEs can be derived from this Lagrangian with standard methods, see, for example, \cite{Roberts:1994dr,Alkofer:2000wg,Alkofer:2008nt}. Since vertex DSEs are already rather complex the program \textit{DoFun} \cite{Alkofer:2008nt,Huber:2011qr} was used for their derivation. 

\begin{figure}[tb]
\includegraphics[width=8.3cm]{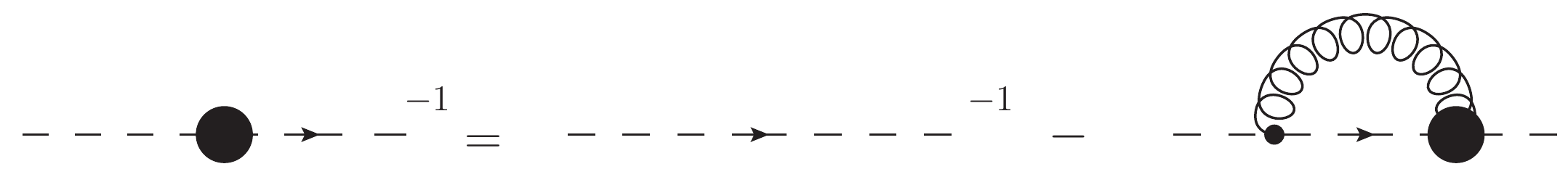}
\vskip5mm
\includegraphics[width=11cm]{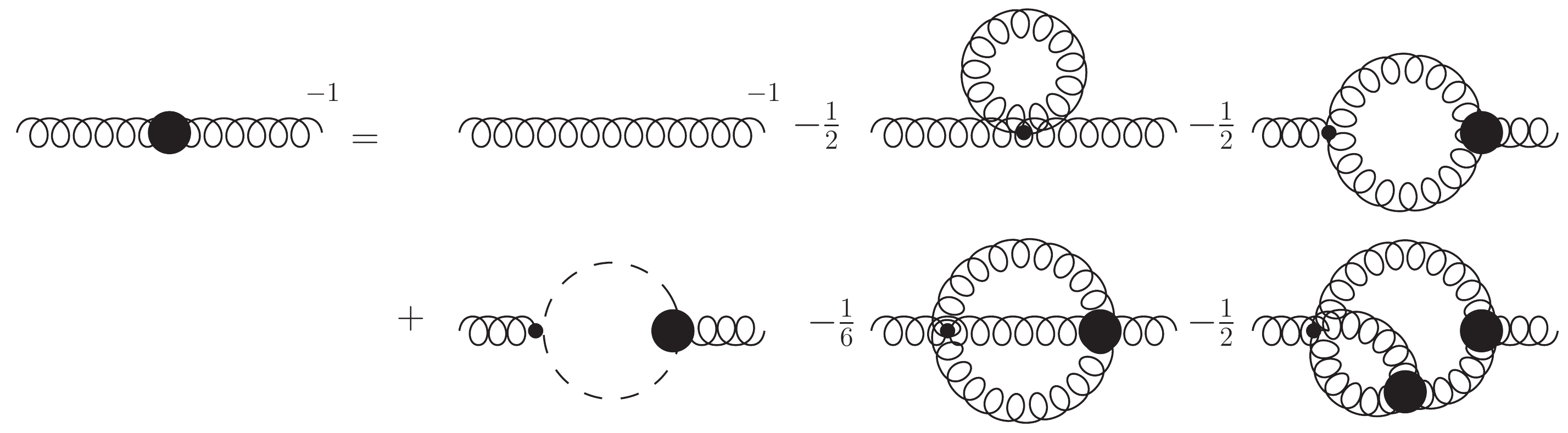}
\begin{center}
\caption{\label{fig:prop-DSEs_2d}Full two-point Dyson-Schwinger equations of Landau gauge Yang-Mills theory. All internal propagators are dressed. Thick blobs denote dressed vertices. Wiggly lines are gluons, dashed ones ghosts.}
\end{center}
\end{figure}

The propagators of the ghost and gluon fields are given by
\begin{align}
 D_{gh}(p^2):=-\frac{G(p^2)}{p^2}, &\quad  D_{gl,\mu\nu}(p^2):=P_{\mu\nu}(p)\frac{Z(p^2)}{p^2},
\end{align}
respectively, where the transverse projector is $P_{\mu\nu}(p)=\delta_{\mu\nu}-p_\mu p_\nu/p^2$. Their full DSEs are given in \fref{fig:prop-DSEs_2d}. In the IR we will use the parametrization
\begin{align}\label{eq:props_power_law}
 G^{IR}(p^2)=B\cdot(p^2)^{\de_{gh}}, &\quad  Z^{IR}(p^2)=A\cdot(p^2)^{\de_{gl}}
\end{align}
for the dressing functions, where $\de_{gh}$ and $\de_{gl}$ are the so-called IR exponents which describe the IR behavior of the dressings qualitatively. For the scaling solution they are related by \cite{Zwanziger:2001kw,Lerche:2002ep}
\begin{align}\label{eq:scal-rel-d}
 2\de_{gh}+\de_{gl}=\frac{(4-d)}{2}.
\end{align}
Consequently, in the case of the scaling solution the IR behavior of both dressing functions can be described by one variable only which we will denote by $\ka:=-\de_{gh}$.

For the two-point DSEs we adopt a standard truncation scheme from four dimensions where all diagrams involving the bare four-gluon vertex are dropped. Note that the four-gluon vertex only appears in the gluon DSE and all conclusions drawn from the ghost DSE in Sec.~\ref{sec:gh-DSE} below are unaffected by our truncation, in particular the conclusion that in two dimensions only the scaling solution exists. We can thus use the available information on the vertices \cite{Huber:2007kc,Alkofer:2008dt} for this solution type to infer that the neglected two-loop diagrams cannot interfere with the IR asymptotics. Actually they are even more IR suppressed than the gluon loop which is itself subleading in the deep IR. Note that such IR considerations can be done for the complete tower of functional equations \cite{Fischer:2009tn,Huber:2009wh} with the only possible caveat being cancellations between IR leading contributions. Furthermore, two-loop diagrams contribute with a higher power of the coupling in the UV and are thus normally negligible there also. However, in the mid-momentum regime they contribute quantitatively to the gluon dressing, which, as we will demonstrate in Sec.~\ref{sec:mid-momentum}, can have an  important impact on the UV behavior of the ghost in two dimensions. For more details we refer to that section.

In order to obtain a scalar equation from the gluon DSE we project it with the transverse projector, yielding
\begin{align}
\label{eq:gh-DSE}
 \frac{1}{G(p^2)}&=1+N_c\,g^2\,\int_q Z(q^2)G((p+q)^2) K_{G}(p,q)\Gamma^{A\bar{c}c}(q;p+q,p),\\
\label{eq:gl-DSE}
 \frac{1}{Z(p^2)}&=1+N_c\,g^2\,\int_q G(q^2)G((p+q)^2) K_{Z}^{gh}(p,q)\Gamma^{A\bar{c}c}(p;p+q,q)\nnnl
 &+N_c\,g^2\,\int_q Z(q^2)Z((p+q)^2) K_{Z}^{gl}(p,q)\Gamma^{A^3}(p,q,p+q).
\end{align}
The quantities $\Gamma^{A\bar{c}c}$ and $\Gamma^{A^3}$ are dressing functions of the ghost-gluon and three-gluon vertices, respectively, and $\int_q$ stands for $\int d^2q/(2\pi)^2$.
The kernels $K_{G}$, $K_Z^{gh}$ and $K_Z^{gl}$ are given explicitly in Appendix~\ref{sec:app_kernels}.

The dressed ghost-gluon vertex is described by two dressing functions
\begin{align}
 \Gamma^{A\bar{c}c,abc}_\mu(k;p,q):=i\,g\,f^{abc}\left(P_{\mu\nu}(k)p_\nu D^{A\bar{c}c}_t(k;p,q)+k_\mu D^{A\bar{c}c}_l(k;p,q)\right).
\end{align}
The basis tensors have been chosen such that $D^{A\bar{c}c}_t(k;p,q)$
and $D^{A\bar{c}c}_l(k;p,q)$ are the purely transverse and
longitudinal dressing functions, respectively. After transverse
projection then only $D^{A\bar{c}c}_t(k;p,q)$ contributes and
$\Gamma^{A\bar{c}c}(k;p,q)$ from eqs.~(\ref{eq:gh-DSE}) and
(\ref{eq:gl-DSE}) can be identified with
$D^{A\bar{c}c}_t(k;p,q)$.\footnote{In the notation of
  \cite{Lerche:2002ep} we have $D^{A\bar{c}c}_t(k;p,q)=A(k;p,q)$ and
  $D^{A\bar{c}c}_l(k;p,q)=B(k;p,q)+A(k;p,q)\,p\cdot k/k^2$.} 
The bare vertex is given by  $i\,g\,f^{abc} p_\mu$. 

The three-gluon vertex has in general 14 Lorentz tensors. Four of them are transverse and thus contribute in the Landau gauge. Furthermore we only consider the color antisymmetric part. As part of our truncation we exclusively take into account the tree-level tensor:
\begin{align}
 \Gamma^{A^3,abc}_{\mu\nu\rho}(p,q,r) = i\,g\,f^{abc}\left(g_{\mu\nu}(q-p)_\rho+g_{\nu\rho}(r-q)_\mu+g_{\rho\mu}(p-r)_\nu\right)D^{A^3}(p,q,r).
\end{align}
Thus $\Gamma^{A^3}(p,q,r)$ in eqs.~(\ref{eq:gh-DSE}) and (\ref{eq:gl-DSE}) is identical to $D^{A^3}(p,q,r)$.

Specific expressions for the dressing functions of the vertices will be given below.

\section{The ghost Dyson-Schwinger equation}
\label{sec:gh-DSE}

This DSE is the simplest one because it has only one integral with a relative simple structure (compared to the gluon loop of the gluon DSE) and is automatically UV finite. Consequently it is in general relatively easy to study numerically using input for the gluon propagator and the ghost-gluon vertex, see, for example, \cite{Boucaud:2008ji,Dudal:2012zx}. In the two-dimensional case the ghost DSE provides additional important information. First we investigate the existence of various types of solutions, then the influence of the mid-momentum regime on the UV behavior of the ghost propagator is scrutinized. In this section we use a bare ghost-gluon vertex and various ans\"atze for the gluon dressing function.

\subsection{Existence of decoupling and scaling solutions}
\label{sec:dec-scal}

In functional equations the choice between decoupling or scaling solutions is realized via the ghost Dyson-Schwinger equation. Employing the subtracted DSE,
\begin{align}\label{eq:sub_gh_DSE}
 \frac{1}{G(p^2)}-\frac{1}{G(p_0^2)}=&\int_q Z(q^2)G((p+q)^2) K_{G}(p,q)\Gamma^{A\bar{c}c}(q;p+q,p)\nnnl
&-\int_q Z(q^2)G((p_0+q)^2) K_{G}(p_0,q)\Gamma^{A\bar{c}c}(q;p_0+q,p_0),
\end{align}
we have to specify the value of the ghost dressing function at $p_0$. Most conveniently one may choose $p_0=0$. For a finite value of $1/G(0)$ a decoupling solution is obtained, whereas the scaling solution corresponds to $1/G(0)=0$. When solving the coupled system of the two-point DSEs the gluon propagator then automatically becomes of the decoupling or scaling type in four dimensions \cite{vonSmekal:1997is,Lerche:2002ep,Fischer:2008uz}.

\begin{figure}[tb]
 \begin{center}
 \begin{minipage}[t]{0.48\textwidth}
  \includegraphics[width=\textwidth]{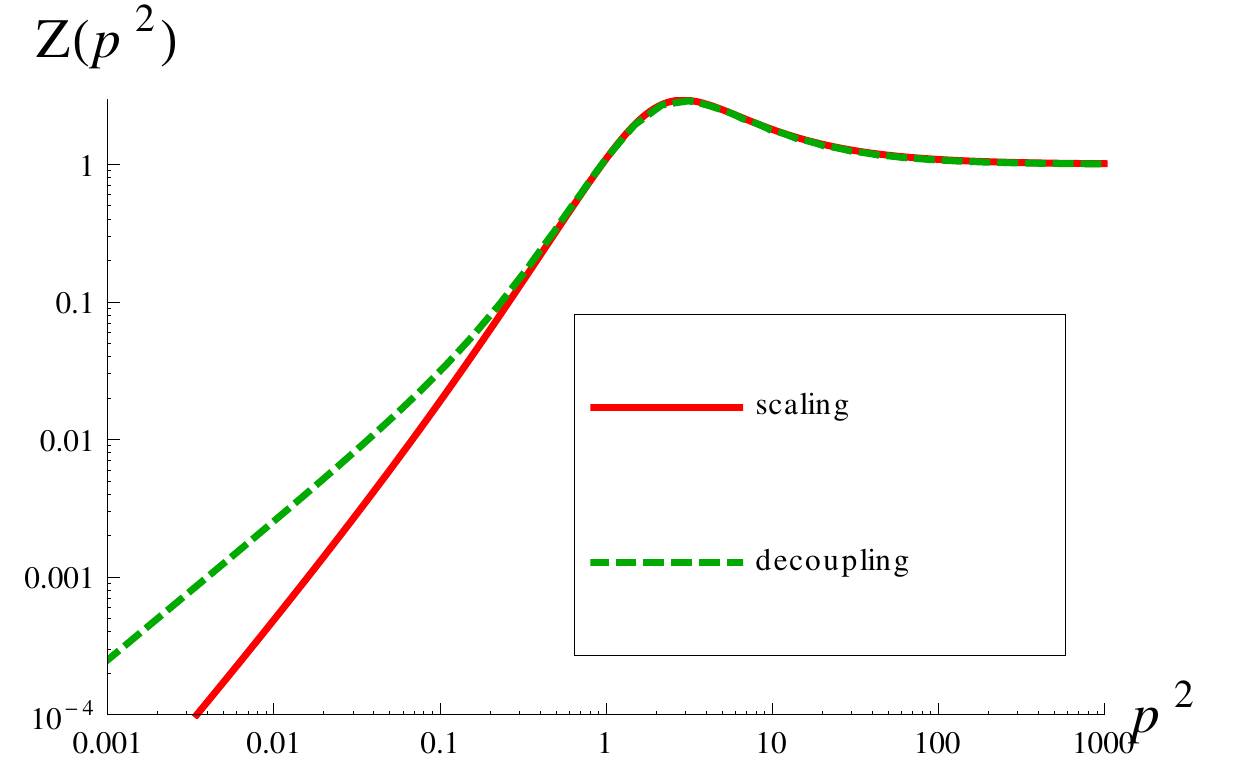}
  \caption{\label{fig:gh_only_gl_ansaetze}Input gluon propagator dressing functions of decoupling and scaling types from \eref{eq:gl-ansatz} with $c_{gl}=1$ and $\ka=1$ (decoupling) and $\ka=1.4$ (scaling). This plot and the following were created with \textit{Mathematica} \cite{Wolfram:2004}.}
 \end{minipage}
 \hfill
 \begin{minipage}[t]{0.48\textwidth}
  \includegraphics[width=\textwidth]{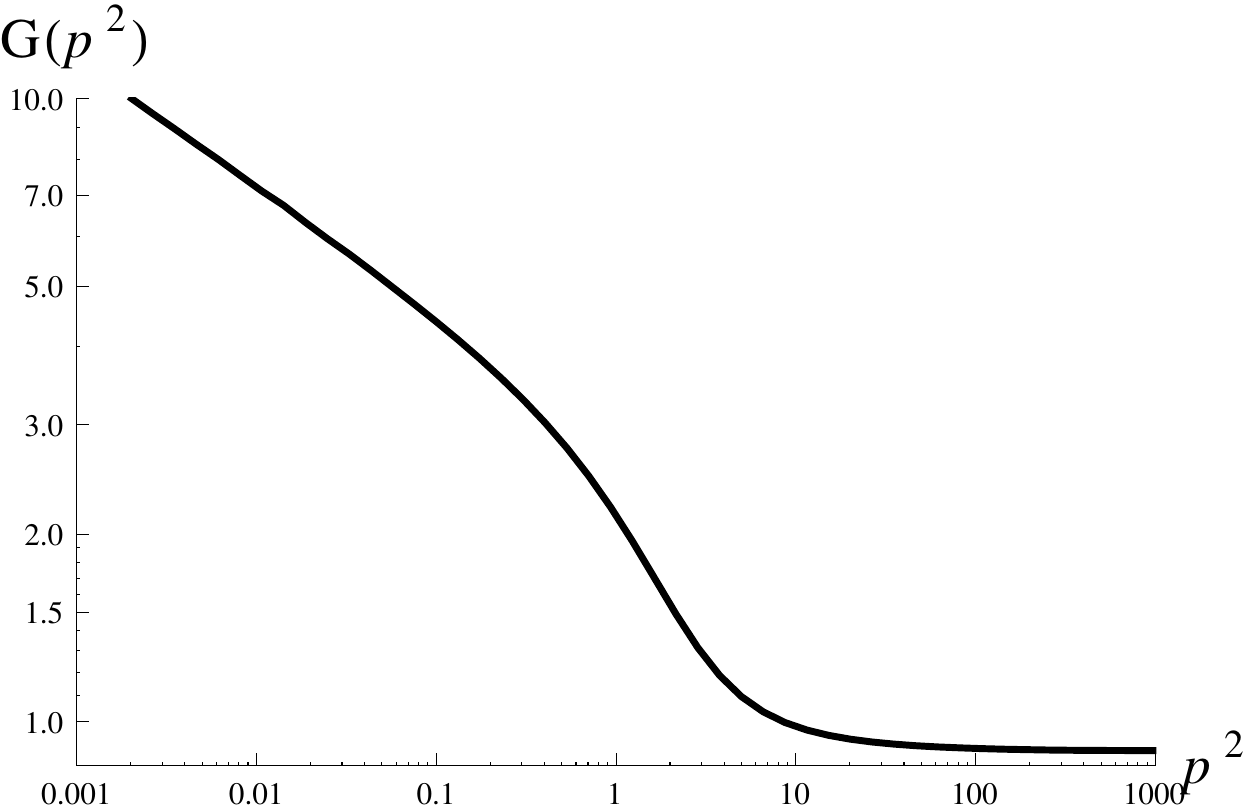} 
  \caption{\label{fig:gh_only_gh_scaling}Ghost propagator dressing function resulting from the scaling type gluon propagator.}
 \end{minipage}
 \begin{minipage}[t]{0.48\textwidth}
  \includegraphics[width=\textwidth]{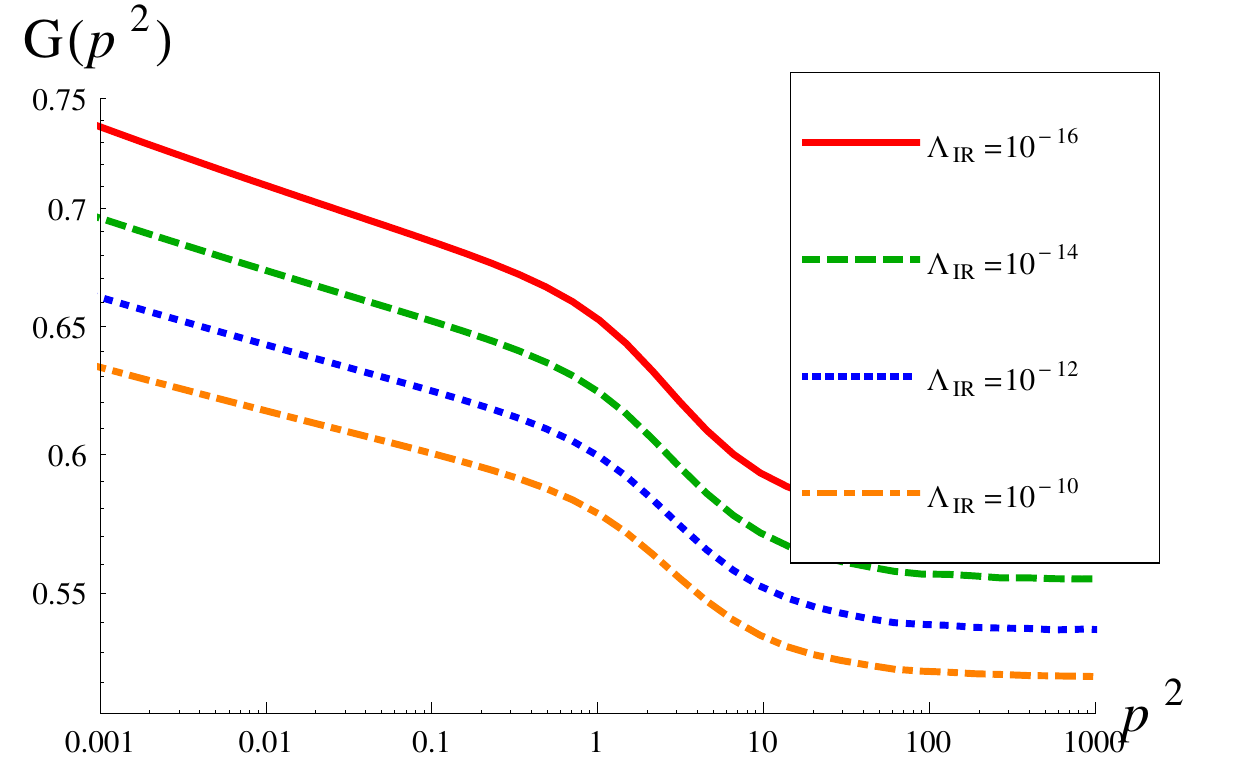}    
  \caption{\label{fig:gh_only_gh_decoupling}Ghost propagator dressing functions resulting from the decoupling type gluon propagator. The various curves correspond to different IR cutoffs.}
 \end{minipage} 
 \end{center}
\end{figure}

The unsubtracted ghost DSE in two dimensions was considered in Ref.~\cite{Cucchieri:2012cb}, where it was found that in order to avoid IR divergences at $p^2=0$ the gluon propagator has to vanish at zero momentum. Here we will corroborate these results numerically. Therefor we use various ans\"atze for the gluon dressing function, which are depicted in \fref{fig:gh_only_gl_ansaetze}:
\begin{align}\label{eq:gl-ansatz}
 Z_{ans}(p^2)=\frac{1}{4} \left(\frac{1}{p^2+1}\right)^2(p^2)^{\de_{gl}}+\left(\frac{p^2}{p^2+1}\right)^2 \frac{1}{1+1/p^2}+\frac{1}{2}c_{gl}\frac{10 \,p^4}{10+p^6}.
\end{align}
For now the parameter $c_{gl}$ is set to one. The IRE exponent $\de_{gl}$ is either $1$ for the decoupling type ansatz or $1.4$ for the scaling type ansatz \cite{Zwanziger:2001kw}.

The subtracted ghost DSE in \eref{eq:sub_gh_DSE} was solved using a standard fixed point iteration.
The resulting ghost propagator dressing functions are shown in figs.~\ref{fig:gh_only_gh_scaling} and \ref{fig:gh_only_gh_decoupling}. For the scaling type ansatz the expected power law is perfectly obeyed in the IR, while for the decoupling type ansatz no valid solution is found. First of all the obtained ghost dressing function, depicted in \fref{fig:gh_only_gh_decoupling}, does not become constant in the IR but diverges. Fitting the IR exponent locally, i.~e., extracting the IR exponent from two adjacent points, shows that the exponent does not settle to a constant value but keeps decreasing slowly. Secondly the ghost dressing depends on the value of the used IR cutoff, as shown in \fref{fig:gh_only_gh_decoupling}, hinting at the occurrence of IR divergences. Trying several variations of the gluon dressing function ansatz could not remove this problem as expected from the analytic considerations of Ref.~\cite{Cucchieri:2012cb}. For the scaling type solution the results for the same IR cutoffs as used for the decoupling type solution lie on top of each other.

Note that both the analytic \cite{Cucchieri:2012cb} and the numeric arguments are based on the assumption that the ghost-gluon vertex is constant in the IR and does not deviate much from tree-level at intermediate momenta. From lattice calculations alone, however, we can strictly speaking not infer this also for a decoupling solution, because what is seen there seems to be a scaling solution. Also the often cited Taylor argument \cite{Taylor:1971ff} does not forbid a ghost-gluon vertex that is IR vanishing. Of course such a behavior is unlikely and unexpected, but one can, in  principle, construct a ghost-gluon vertex that allows a solution for the ghost equation also in this case. We also tried the opposite way, namely to solve the ghost-gluon vertex DSE using decoupling type ans\"atze for the propagators. However, in this case we could not find a solution. This further corroborates the non-existence of a decoupling type solution in two dimensions.

\subsection{Influence of the mid-momentum regime on the ghost's UV behavior}
\label{sec:mid-momentum}

\begin{figure}[tb]
 \begin{center}
 \begin{minipage}[t]{0.48\textwidth}
  \includegraphics[width=\textwidth]{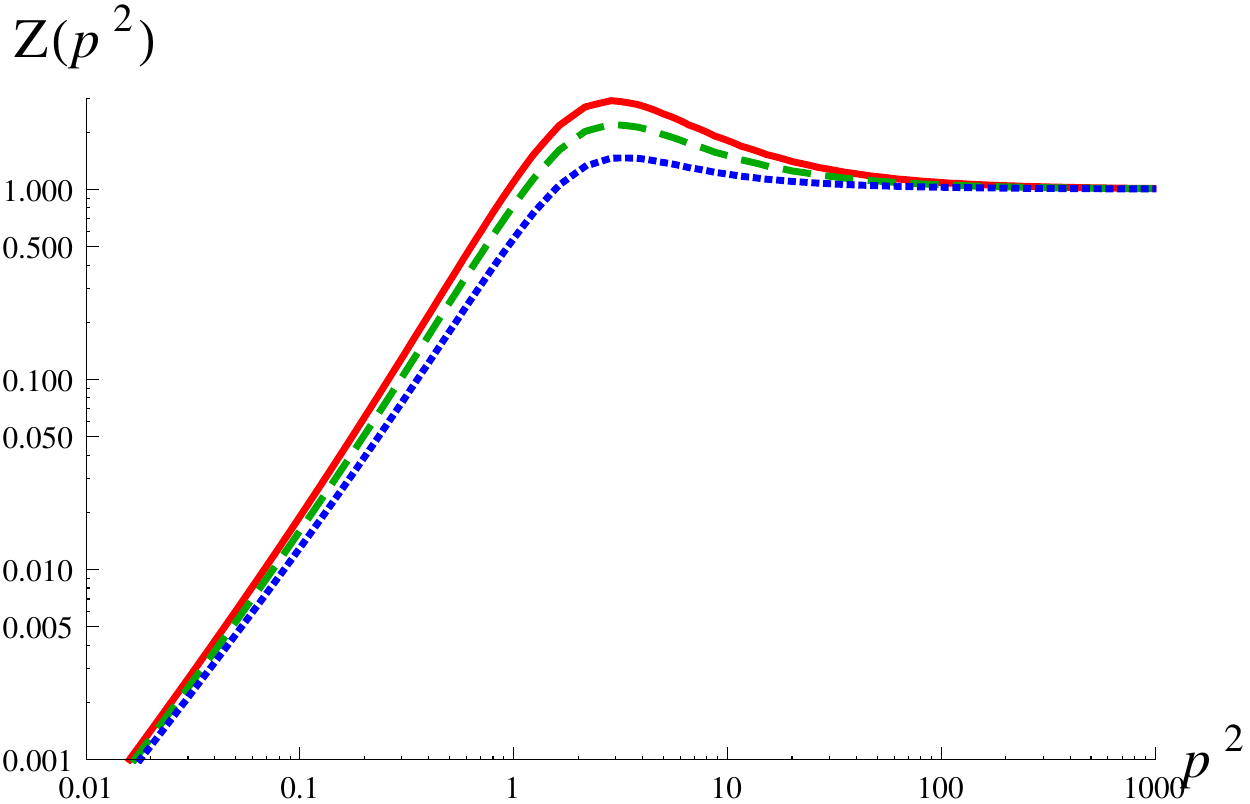}
  \caption{\label{fig:gh_only_gl_midMom_gl}Input gluon propagator dressing functions given by \eref{eq:gl-ansatz} and $c_{gl}=1,\,0.7, \,0.4$ (solid/red, dashed/green, dotted/blue).}
 \end{minipage}
 \hfill
 \begin{minipage}[t]{0.48\textwidth}
  \includegraphics[width=\textwidth]{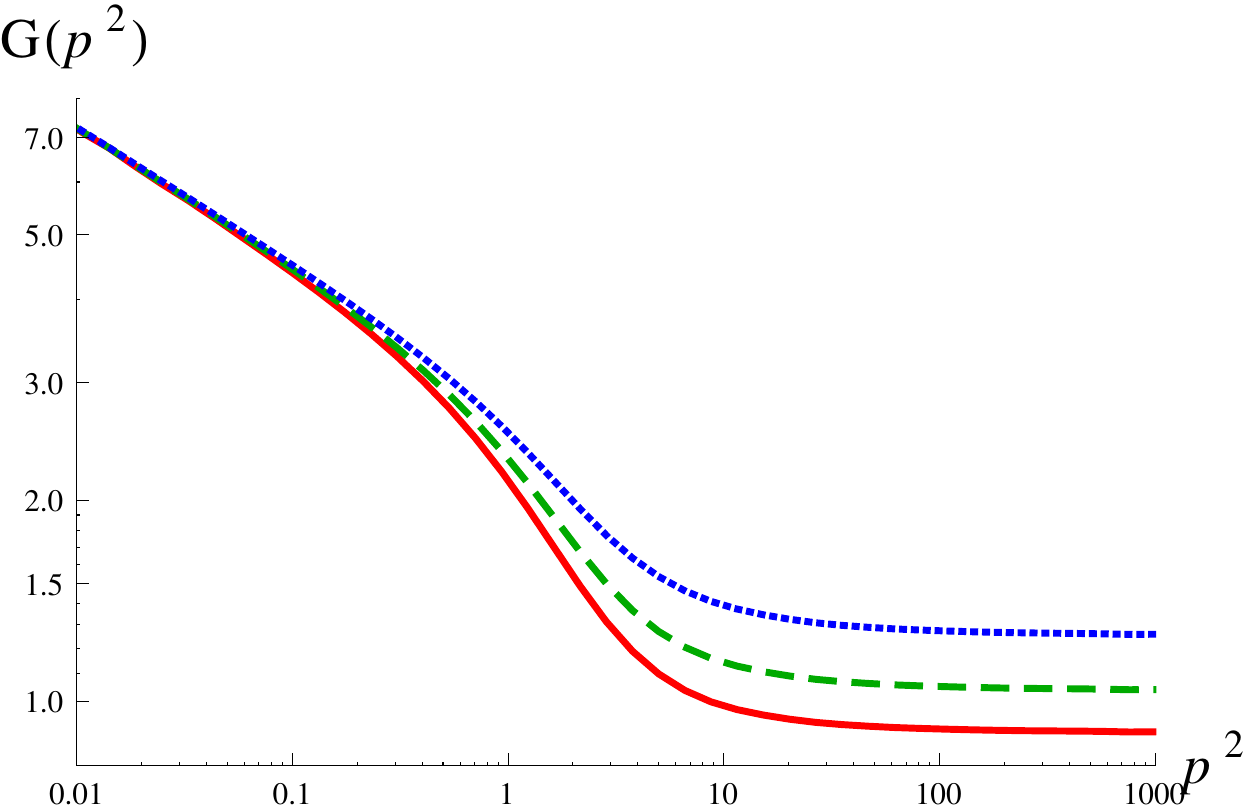} 
  \caption{\label{fig:gh_only_gh_midMom_gh}Ghost propagator dressing functions resulting from the gluon propagator ans\"atze given in \fref{fig:gh_only_gl_midMom_gl} and a bare ghost-gluon vertex. Solid/red, dashed/green and dotted/blue lines correspond to $c_{gl}=1,\,0.7, \,0.4$.}
 \end{minipage}
\end{center}
\end{figure}

Another important lesson from the ghost DSE alone concerns the influence of different momentum regions on each other. In four dimensions such an influence is subleading, mainly due to renormalization. In three dimensions, this is a weak quantitative effect \cite{Maas:2004se}. For example, the UV behavior can be determined self-consistently. In two dimensions this is not the case. This can already be inferred from a purely perturbative investigation based on bare propagators which fails because of IR divergences which can only be remedied by taking into account a non-trivial behavior at momenta below the UV regime. We explicitly demonstrate the influence of the mid-momentum regime by studying the ghost DSE with the gluon dressing function given in \eref{eq:gl-ansatz} where we vary the parameter $c_{gl}$ which modifies the height of the bump in the gluon dressing function. The ans\"atze for $c_{gl}=0.4,\,0.7,\,1$ are shown in \fref{fig:gh_only_gl_midMom_gl} and the resulting ghost dressing functions in \fref{fig:gh_only_gh_midMom_gh}. It is clearly visible that the height of the bump is correlated with the value of the ghost dressing function in the UV: The higher the bump the lower the ghost dressing becomes in the UV. This observation will be important later in Sec.~\ref{sec:gh+gl} when we solve the DSEs of both propagators simultaneously. Note that the IR is not affected by the mid-momentum behavior. 

\section{The IR solutions revisited: Analytic considerations}
\label{sec:IR-anal}

\subsection{The solution $\ka=0$}

For the scaling type solution one can determine the qualitative behavior in the deep IR by calculating the IR exponents which can be done analytically \cite{vonSmekal:1997vx,Zwanziger:2001kw,Lerche:2002ep}. Therefor one calculates the IR leading diagrams in the two-point DSEs, viz.\ the ghost loop in the gluon DSE and the loop of the ghost DSE, see \fref{fig:prop-DSEs_2d}, using the power law ans\"atze for the propagator dressing functions given in \eref{eq:props_power_law} together with the scaling relation \eref{eq:scal-rel-d}:
\begin{align}
 Z^{IR}(p^2)=A\cdot (p^2)^{2\ka+(4-d)/2}, \quad G^{IR}(p^2)=B\cdot (p^2)^{-\ka}.
\end{align}
For the ghost-gluon vertex usually a bare vertex is employed but the same results for $\kappa$ hold for all ghost-gluon vertices with a regular IR limit \cite{Lerche:2002ep}. The IR value of the coupling, however, depends on the IR value of the vertex.
The calculation of $\kappa$ in $d$ dimensions boils down to solving the following equation \cite{Zwanziger:2001kw,Lerche:2002ep}:
\begin{align}\label{eq:kappa_eq}
 \frac{\sin(\pi \,\ka)\Gamma(d/2-\ka)\Gamma(\ka)\Gamma(1+d/2+\ka)}{2(d-1)\sin(\pi(d/2-2\ka))\Gamma(d-2\ka)\Gamma(2\ka)\Gamma(1+\ka)}=1.
\end{align}

For two dimensions two solutions have been cited in
Ref.~\cite{Zwanziger:2001kw}: $\ka_1=0.2$, which is close to values
extracted from lattice data
\cite{Maas:2007uv,Maas:2009ph,Cucchieri:2011ig}, and $\ka_2=0$. Note
that the IR exponents of the propagators for the second solution are
$\de_{gh}=0$ and $\de_{gl}=1$ which are exactly the IR exponents of
the decoupling type solutions. Hence the two-dimensional case is
special insofar as a possible decoupling solution would also respect the
scaling relation \eref{eq:scal-rel-d}. We will now present further
evidence against the decoupling type solution with $\ka_2=0$ in
addition to those presented in \cite{Cucchieri:2012cb}.

\begin{figure}[tb]
\begin{center}
 \includegraphics[width=0.6\textwidth]{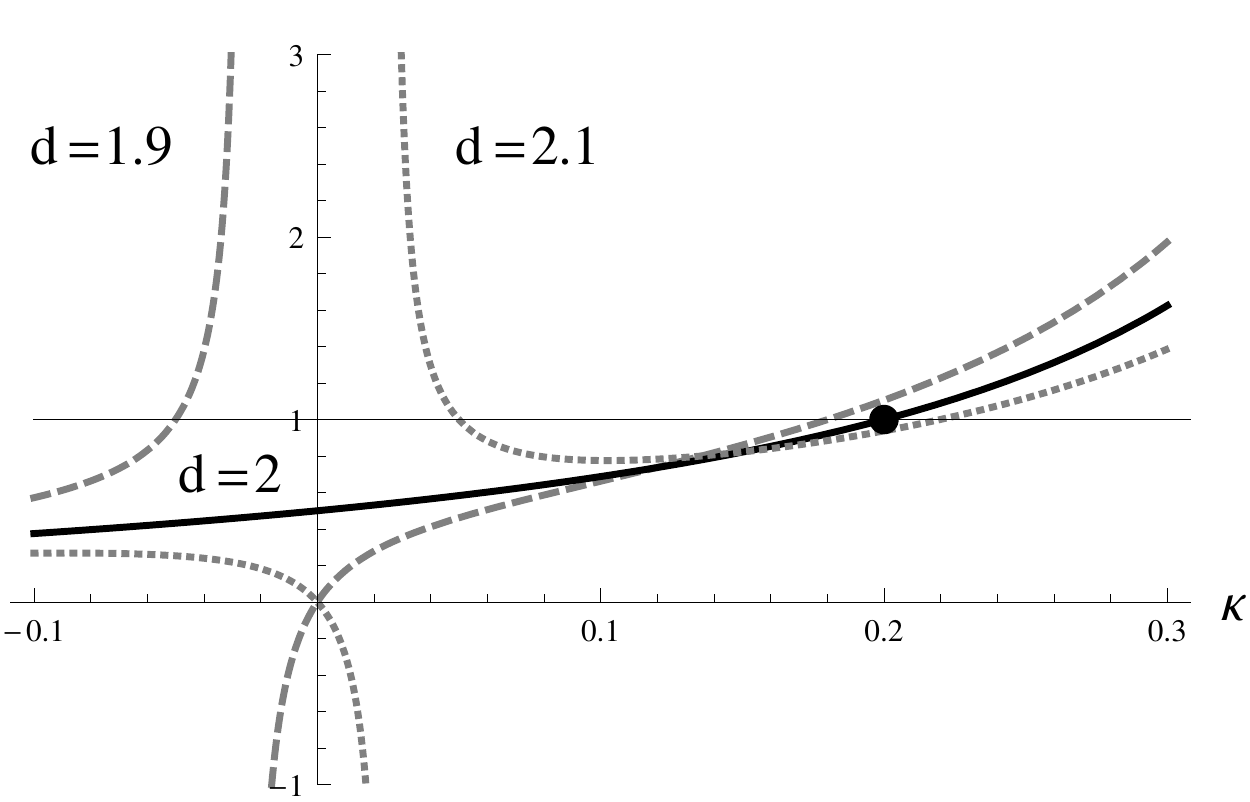}
\caption{\label{fig:kappa_0}Left- and right hand sides of \eref{eq:kappa_eq}. The dashed, solid and dotted lines corresponds to $d=1.9,\,2,\,2.1$, respectively. As $d$ approaches $2$, a solution seems to be $\ka=0$, but actually the point $d=2$, $\ka=0$ is multi-valued. The blob at $\ka=0.2$ represents the standard solution.}
\end{center}
\end{figure}

Let us look at \eref{eq:kappa_eq} for $d=2$ and $\kappa=0$. In
Ref. \cite{Zwanziger:2001kw} it was argued that for $d=2+\epsilon$ a
solution $\ka=0+\epsilon$ exists. This is illustrated in
\fref{fig:kappa_0}, where the left-hand side of \eref{eq:scal-rel-d}
is plotted for $d=1.9,\,2,\,2.1$. As $d$ approaches $2$ from either
side, there are intersection points with unity which 
indeed approach $\ka=0$, such that there seems
to be a solution also for $d=2$.  On the other hand, we can take the
limit $d\rightarrow 2$ on the left-hand side analytically: 
\begin{align}
 \frac{1+\ka}{2-4\ka}=1.
\end{align}
The only solution to this is $\ka=0.2$. The reason for this discrepancy is that $d=2$ and $\ka=0$ corresponds to a multi-valued point and a wide range of possible values can be obtained. For example, we could approach the point by taking the limit $\epsilon\rightarrow 0$ and $d=2+2\epsilon$ and $\ka=\epsilon$. Then the left-hand side of \eref{eq:kappa_eq} is one and $\ka=0$ is a solution. Note that in four dimensions the same arguments apply to the solution $\ka=1$. In three dimensions the mathematical existence of two solutions is unequivocal. Varying the number of dimensions continuously one sees that there are two branches of solutions \cite{Zwanziger:2001kw,Maas:2004se}. In four dimensions only one of those two solutions is found to be realized in numeric calculations \cite{Fischer:2002eq}, but in three dimensions solutions for both cases have been found \cite{Maas:2004se}. Nonetheless, one may speculate that only one branch is physical, namely the one for which $\ka=0.595353$ in four and $\ka=0.2$ in two dimensions. The presence of the second solution in three dimensions could be a truncation artifact, but this requires further investigation.

In summary, the infrared exponent $\ka_2=0$ requires an additional prescription
for how to define a two-dimensional solution in a particular 
limit $d\to 2$ suitably combined with $\ka(d) \to 0$. This makes its
existence scheme dependent and it seems at least questionable whether
such a solution is realized. Since this solution coincides with the
decoupling solution, such a type of solution seems, from this point of
view, unlikely to exist in two dimensions. 

\subsection{Scaling solutions with $\ka\neq0.2$}
\label{sec:other_kappa}

Although lattice calculations seem to obtain a scaling type solution,
it is not yet settled what the numerical value of $\ka$ is. First
results corresponded to values of about $0.15$ \cite{Maas:2007uv}, but
recent calculations on larger lattices seem to favor a value of
$0.225$ \cite{Cucchieri:2011ig}. From functional equations, however,
the value $\ka=0.2$ is well established
\cite{Zwanziger:2001kw,Lerche:2002ep}. In the calculation of $\ka$
only the IR parts of the propagators and the ghost-gluon vertex
enter. Hence modifications in the mid-momentum and UV regime are
irrelevant as are any details of the three-gluon vertex as long as it
is not more IR divergent than $(p^2)^{-3\ka+(d-4)/2}$
\cite{Alkofer:2004it,Huber:2007kc}. As far as the propagators are
concerned only the combination $A\,B^2$ of the coefficients of the IR
power laws from \eref{eq:props_power_law} enters and is calculated
together with $\ka$. The only quantity which has to be set by hand is
the ghost-gluon vertex in the IR. A class of regular vertex dressings
was investigated in Ref.~\cite{Lerche:2002ep}, where it was shown that
they all result in the same $\ka$. Regularity means here that the
ghost-gluon vertex dressing function has a unique IR limit when all
momenta go to zero. Thus the only way to obtain a different value for
$\kappa$ is to allow non-regular vertex dressings, e.~g., when the IR limit
depends on the angle between two momenta.  

To implement such an angle dependence we employ the ansatz
\begin{align}\label{eq:ghg_ansatz_angle}
  D^{A\bar{c}c}_{t,ad}(k;p,q)=1+a\,(\cos \varphi)^2
\end{align}
for the transverse dressing function of the ghost-gluon vertex, where $\varphi$ is the angle between the two ghost momenta:
\begin{align}\label{eq:phi}
\cos(\varphi)=\frac{p\cdot q}{|p||q|}=\frac{k^2-p^2-q^2}{2|p||q|}.
\end{align}
The angle is not well defined in the zero-momentum limit.

\begin{figure}[tb]
\begin{center}
 \begin{minipage}[t]{0.48\textwidth}
 \includegraphics[width=\textwidth]{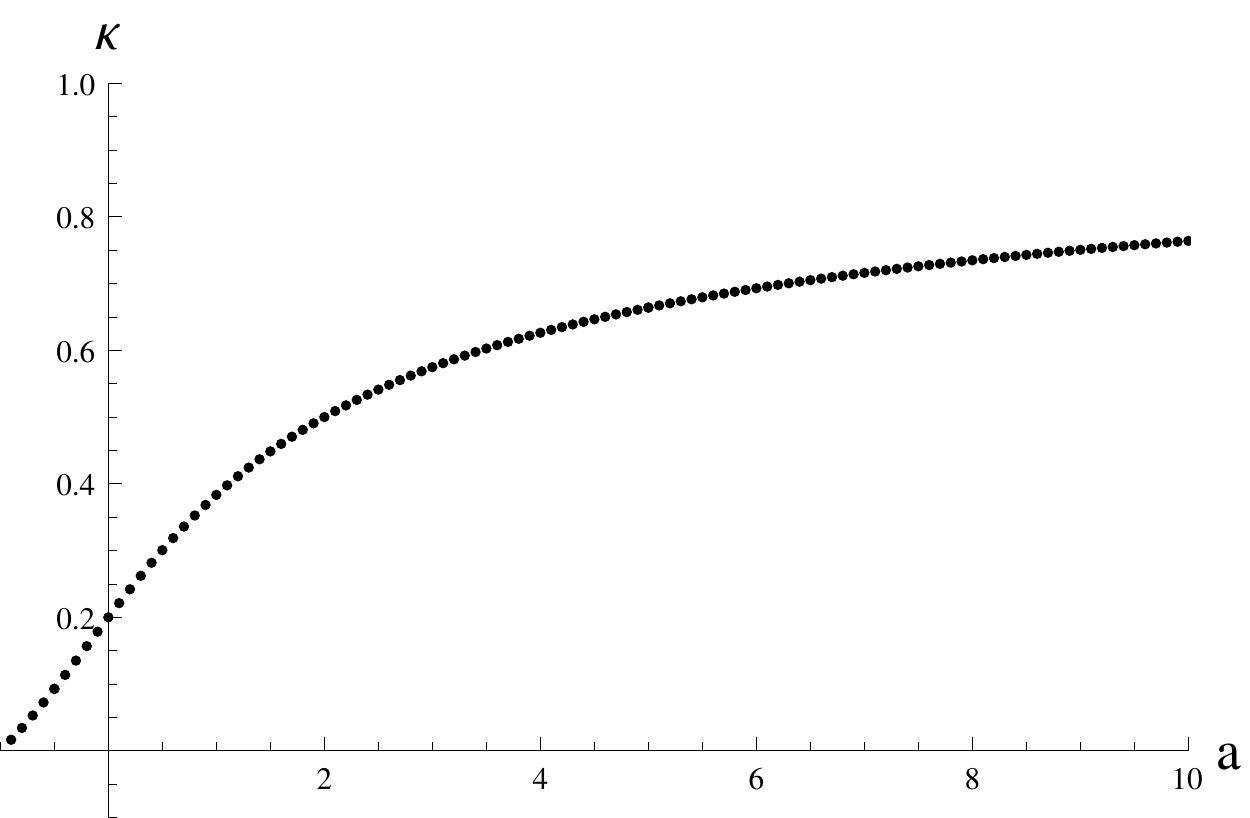}
 \end{minipage}
 \hfill
 \begin{minipage}[t]{0.48\textwidth}
 \includegraphics[width=\textwidth]{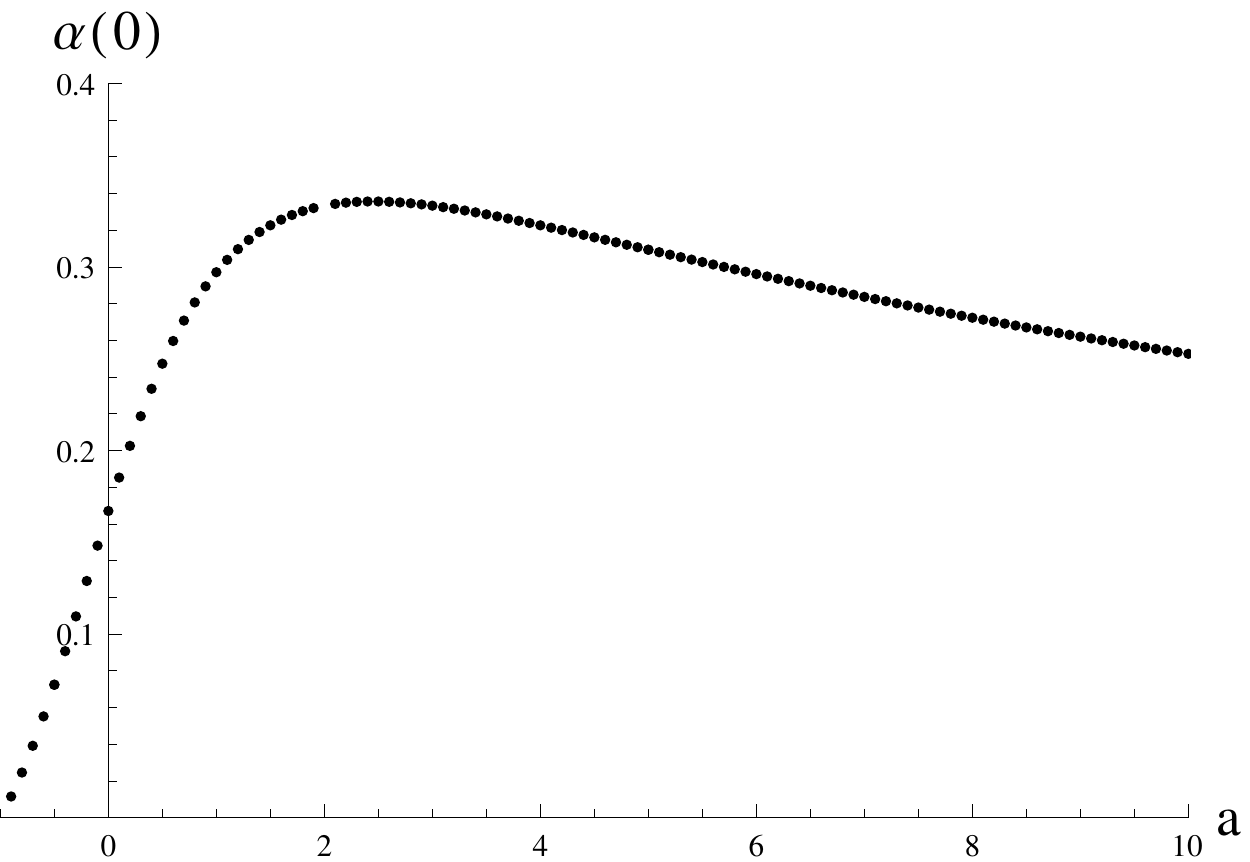}
 \end{minipage}
\caption{\label{fig:a_kappa_alpha0}\textit{Left:} $\ka$ for different values of $a$. \textit{Right:} The corresponding values of $\alpha(0)$. Both plots are for two dimensions.}
\end{center}
\end{figure}

\begin{figure}[tb]
\begin{center}
 \begin{minipage}[t]{0.48\textwidth}
 \includegraphics[width=\textwidth]{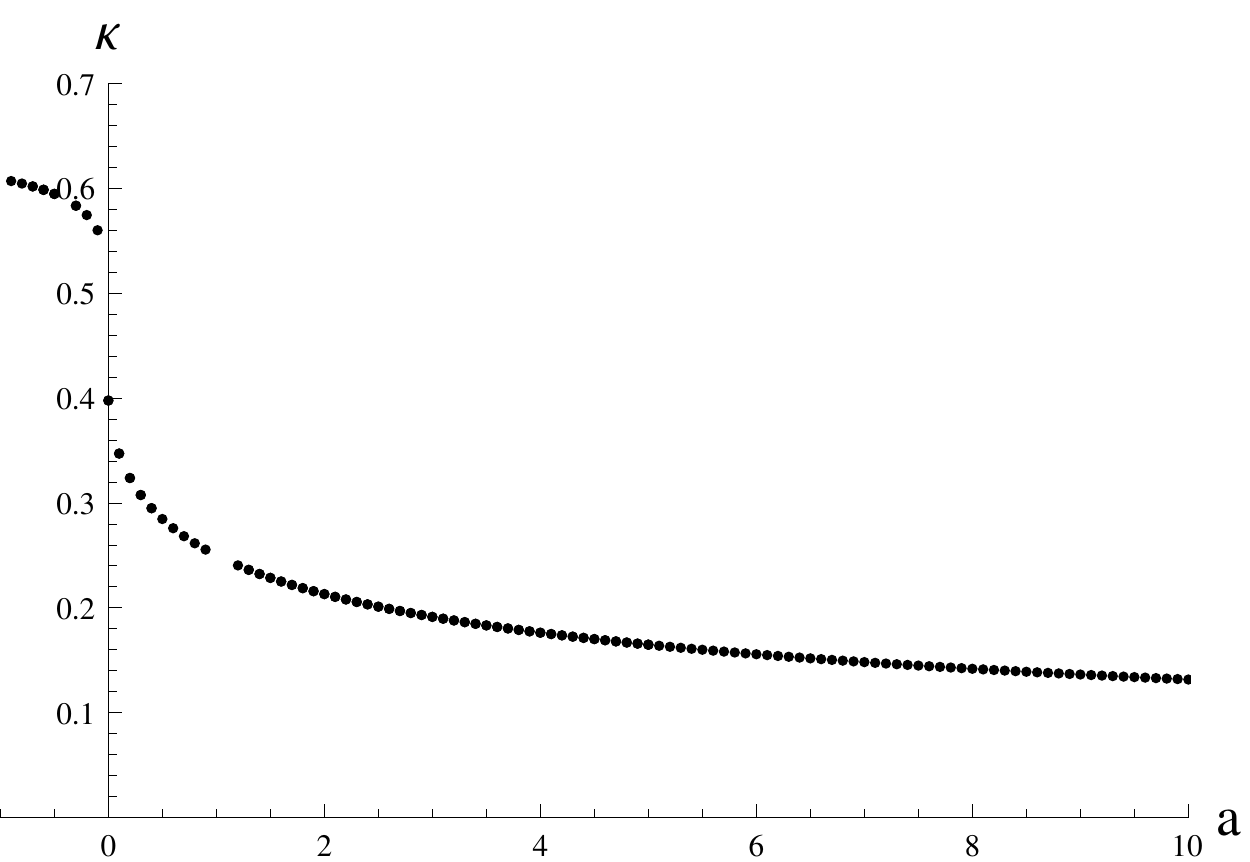}
 \end{minipage}
 \hfill
 \begin{minipage}[t]{0.48\textwidth}
 \includegraphics[width=\textwidth]{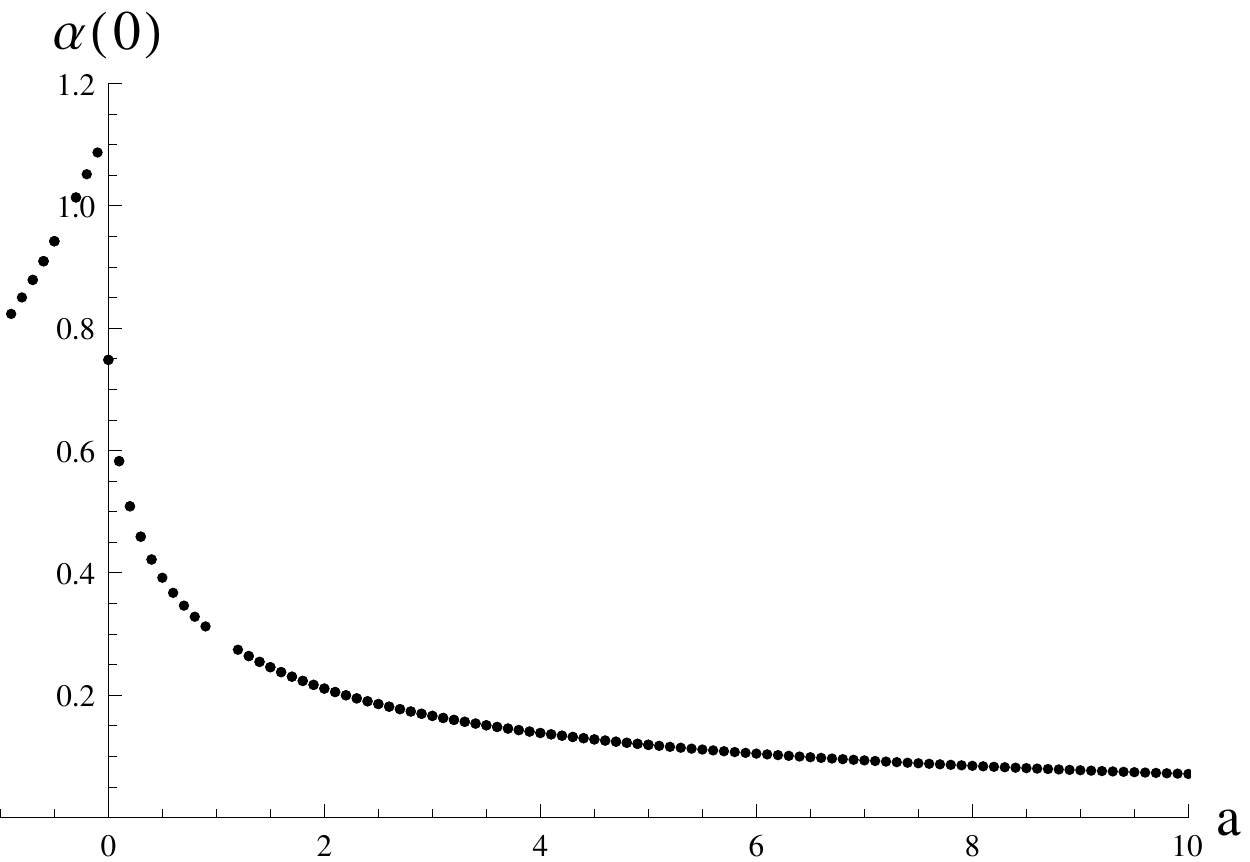}
 \end{minipage}
\caption{\label{fig:a_kappa_alpha0-3d}\textit{Left:} $\ka$ for different values of $a$. \textit{Right:} The corresponding values of $\alpha(0)$. Both plots are for three dimensions and only one solution branch is shown.}
\end{center}
\end{figure}

\begin{figure}[tb]
\begin{center}
 \begin{minipage}[t]{0.48\textwidth}
 \includegraphics[width=\textwidth]{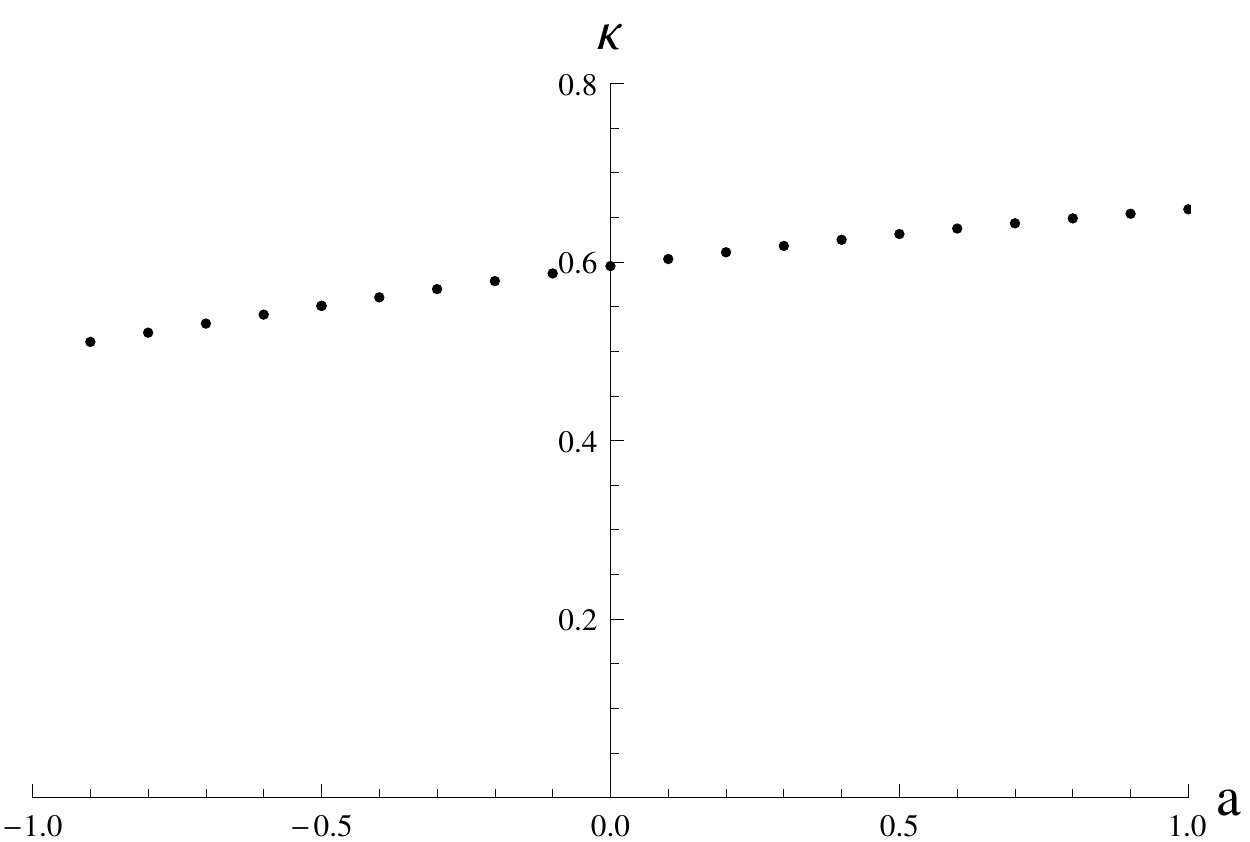}
 \end{minipage}
 \hfill
 \begin{minipage}[t]{0.48\textwidth}
 \includegraphics[width=\textwidth]{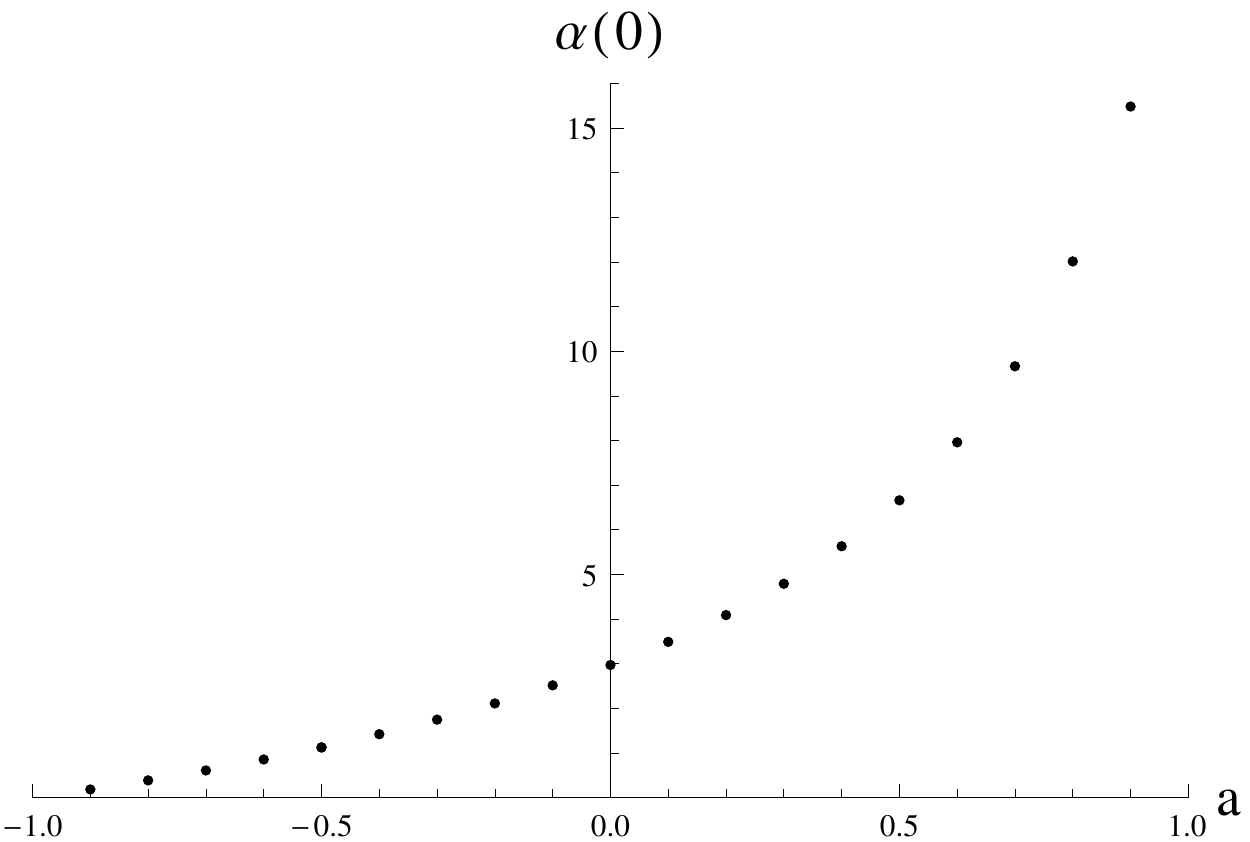}
 \end{minipage}
\caption{\label{fig:a_kappa_alpha0-4d}\textit{Left:} $\ka$ for different values of $a$. \textit{Right:} The corresponding values of $\alpha(0)$. For $a\gtrsim1.3$ the coupling becomes negative and is not shown. Both plots are for four dimensions.}
\end{center}
\end{figure}

The ansatz (\ref{eq:ghg_ansatz_angle}) is motivated by the
ghost-antighost symmetry in Landau gauge and corresponds to the first
two terms in a Fourier series of more generally possible angle
dependencies \cite{Alkofer:2000wg}. The values for $\ka$ and the IR
fixed-point value for the coupling, given by
\cite{vonSmekal:1997is,Lerche:2002ep,Huber:2007kc,vonSmekal:2009ae} 
\begin{align}
 \alpha(0):=\frac{g^2}{4\pi} A\,B^2,
\end{align}
can then be calculated using this ansatz. The results are depicted in \fref{fig:a_kappa_alpha0}. Figs.\ \ref{fig:a_kappa_alpha0-3d} and \ref{fig:a_kappa_alpha0-4d} show results for three and four dimensions, respectively, where the same arguments as in two dimensions apply.

Finally we want to make a comment on the dependence of $\kappa$ on the employed projection of the gluon DSE. In principle any projection should yield the same value for $\kappa$, but since we have to use a truncated system of equations in actual fact it does depend on the projection operator. For example, the Brown-Pennington projector leads to $\ka=0.224745$ in two dimensions. However, any non-transverse projection leads to contributions from the longitudinal part of the ghost-gluon vertex. If we had the exact ghost-gluon vertex, this would be fine and we would get the same value of $\ka$ for all projections. Since we do not have the exact vertex, we will employ the transverse projection only. This disposes of the following ambiguity as shown in Ref. \cite{Lerche:2002ep}: In general there exists a second gauge parameter which interpolates between the normal Faddeev-Popov action of linear covariant gauges and the ghost anti-ghost symmetric gauges. However, in the Landau gauge this parameter is irrelevant and thus any quantity depending on it is ambiguous. A transverse projector gets rid of all such terms.
See also Ref. \cite{Fischer:2010is} for a short discussion of this issue.

\section{The coupled system of two-point DSEs}
\label{sec:gh+gl}

As a next step we solve the DSEs for the ghost and gluon two-point functions simultaneously. This system contains as a new quantity the three-gluon vertex for which we need an ansatz. An expression similar to the one used in four dimensions \cite{Fischer:2002eq,Fischer:2003zc} turns out to be insufficient, because the gluon loop will then dominate the ghost loop at intermediate momenta and the gluon dressing function turns negative. This was already observed in three dimensions \cite{Maas:2004se} and motivated another construction for the three-gluon vertex. A similar one for two dimensions is 
\begin{align}
\label{eq:3g-ans-3d}
 D^{A^3}_{ans}(p,q,r)=\left(\frac{(G(p^2)G(q^2)G(k^2))^{-2-1/\kappa}}{Z(p^2)Z(q^2)Z(k^2)}\right)^\alpha.
\end{align}
This expression becomes $1$ for large momenta and is suppressed for low ones. The exponent $\alpha$ controls the strength of the suppression. In order to find a solution it must not be too small. However, it is known that the three-gluon vertex of the scaling solution is not IR suppressed but actually IR enhanced in two \cite{Maas:2007uv,Huber:2007kc}, three \cite{Huber:2007kc} and four dimensions \cite{Alkofer:2004it}. In two dimensions, an approximate agreement between the prediction from functional methods \cite{Huber:2007kc} and the lattice results \cite{Maas:2007uv} is found, while in higher dimensions the lattice results \cite{Cucchieri:2008qm,Alles:1996ka} do not yet penetrate very far into the infrared. Furthermore, lattice calculations showed that the three-gluon vertex changes its sign at intermediate momenta \cite{Maas:2007uv,Cucchieri:2008qm}. For two dimensions we show in Appendix~\ref{sec:tgvert} that the DSE of the three-gluon vertex reproduces this zero crossing. Based on this behavior we adopt also the following ansatz for the dressing function of the three-gluon vertex:
\begin{align}\label{eq:tg_ansatz}
 D^{A^3}_{mod}(p,q,r)=\frac{p^2+q^2+r^2}{p^2+q^2+r^2+h_{IM}\Lambda^2}+h_{IR}(p^2+q^2+r^2)^{-3\ka-1}.
\end{align}
The first term ensures the correct UV behavior and the second term implements the IR divergence \cite{Huber:2007kc}. Note that the parameter $h_{IR}$ has to be negative in order to reproduce the zero crossing. $h_{IR}$ can be obtained by a fit to lattice data. However, in order to obtain a solution we have to choose a much higher absolute value, probably in order to compensate for dropping the two-loop diagrams in our truncation.

\begin{figure}[tb]
 \begin{center}
 \begin{minipage}[t]{0.48\textwidth}
 \includegraphics[width=\textwidth]{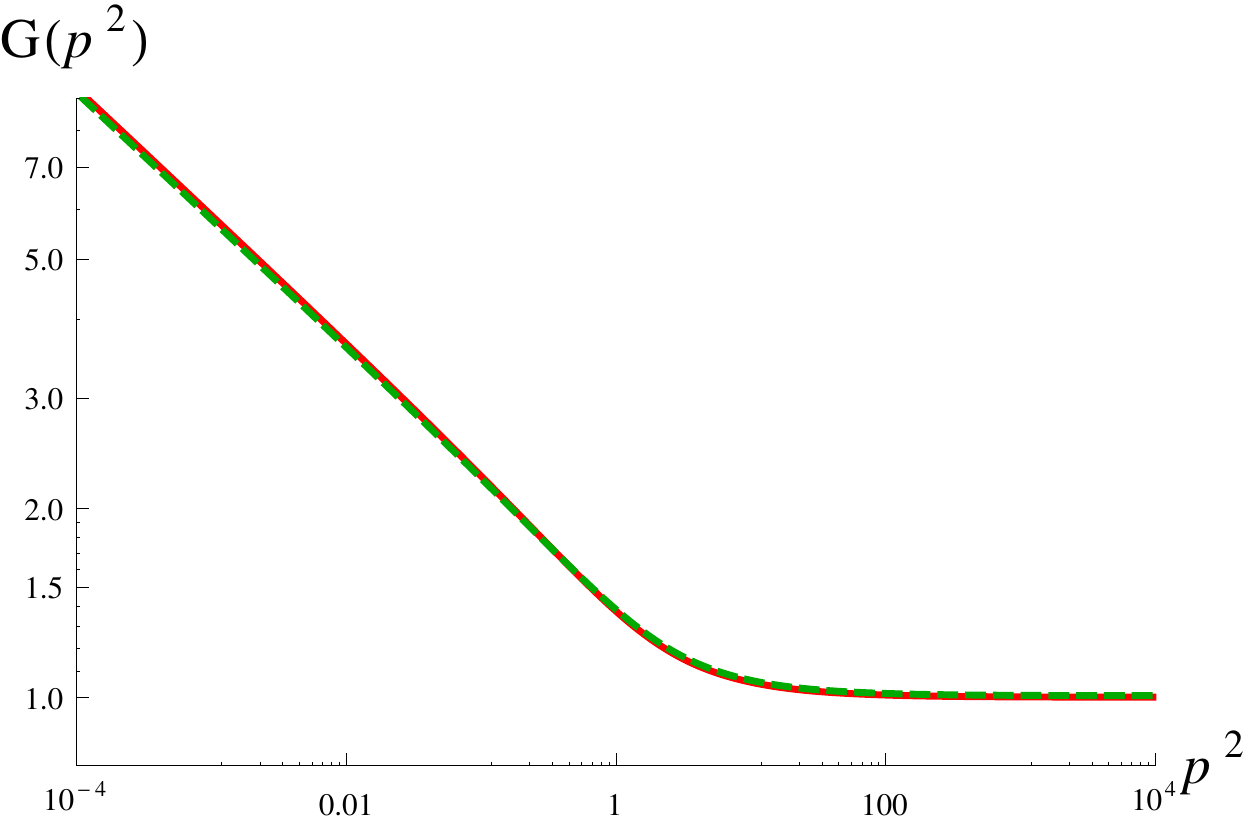}
 \end{minipage}
 \hfill
 \begin{minipage}[t]{0.48\textwidth}
 \includegraphics[width=\textwidth]{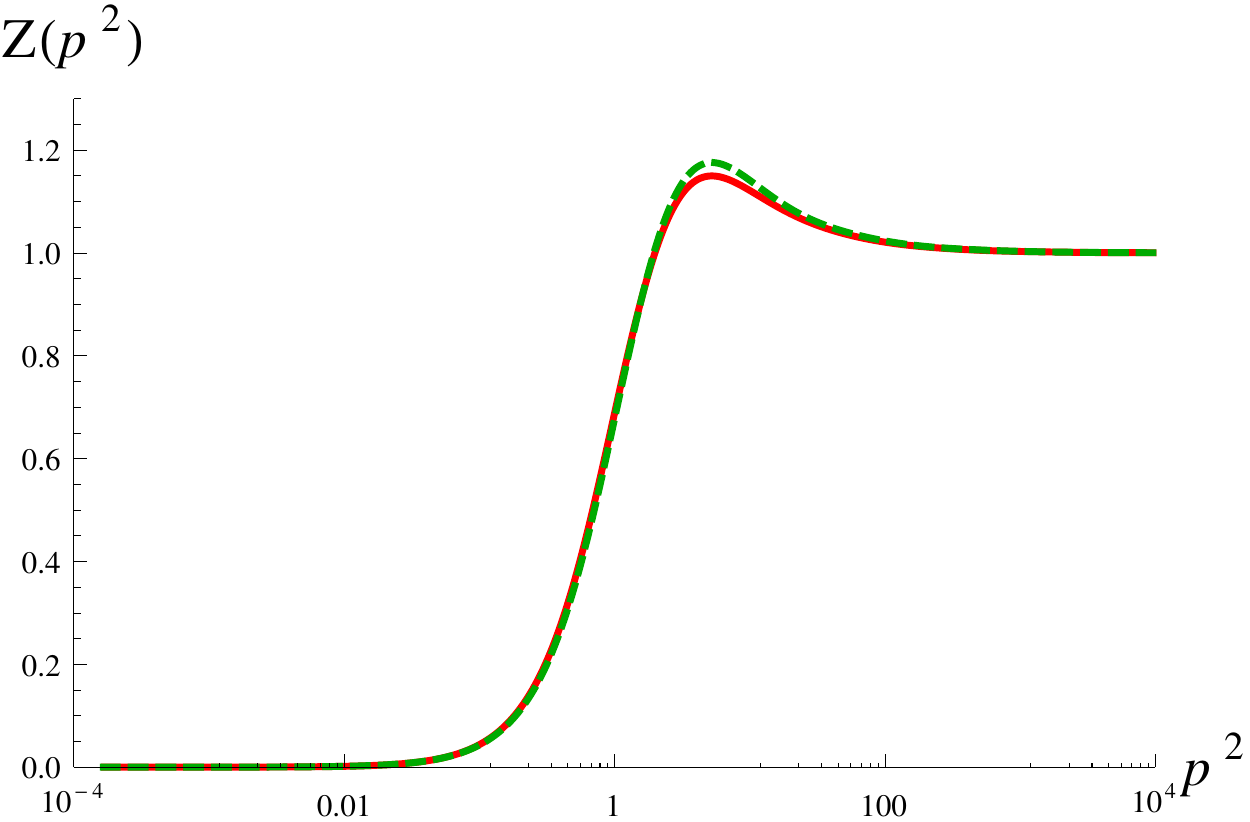}
 \end{minipage}
 \caption{\label{fig:props_VModels_comp}Ghost (\textit{left}) and gluon (\textit{right}) dressings obtained with a bare ghost-gluon vertex and the three-gluon vertex of \eref{eq:3g-ans-3d} with $\alpha=0.09$. The red/solid lines result from subtracting the logarithmic divergences in the gluon loop kernel, while green/dashed lines stem from using the subtraction \eref{eq:Z_subtracted}.}
 \end{center}
\end{figure}

As in four dimensions the gluon DSE contains spurious UV divergences. They are the remnants of the quadratic divergences in four dimensions which arise due to the use of a UV cutoff as regulator and thus appear already at the perturbative level \cite{Maas:2011se}.
One possibility to handle these logarithmic divergences is to modify the integrand of the gluon loop such that the spurious divergences cancel \cite{Fischer:2002eq,Fischer:2003zc}. Another possibility is to employ a counter term, akin to the minimal subtraction scheme from perturbation theory \cite{Maas:2004se,Maas:2011se}. The divergences appear in the gluon two-point DSE in the following form:
\begin{align}
\frac{1}{Z(p^2)}=1+\Sigma(p^2)=1+\frac{g^2}{p^2}\Sigma'(p^2)+c\,g^2\frac{\ln\Lambda^2}{p^2},
\end{align}
where $\Sigma(p^2)$ is logarithmically divergent and $\Sigma'(p^2)$ is finite.
Subtracting from this the self-energy at the momentum $s$ times $s^2/p^2$ yields
\begin{align}\label{eq:Z_subtracted}
 \frac{1}{Z(p^2)}-\frac{1}{Z(s^2)}\frac{s^2}{p^2}=\Sigma(p^2)-\Sigma(s^2)\frac{s^2}{p^2}=\frac{g^2}{p^2}\Sigma'(p^2)-\frac{g^2}{p^2}\Sigma'(s^2),
\end{align}
i.~e., the logarithmic divergence is gone. Since the logarithmic divergences only reside in the part proportional to $p_\mu p_\nu$, its advantageous to use the longitudinally projected self-energy for the subtraction term since this interferes least with the physical part.  The two methods for subtracting the logarithmic divergences are compared in \fref{fig:props_VModels_comp}. 

\subsection{Bare ghost-gluon vertex}

\begin{figure}[tb]
 \begin{center}
 \begin{minipage}[t]{0.48\textwidth}
 \includegraphics[width=\textwidth]{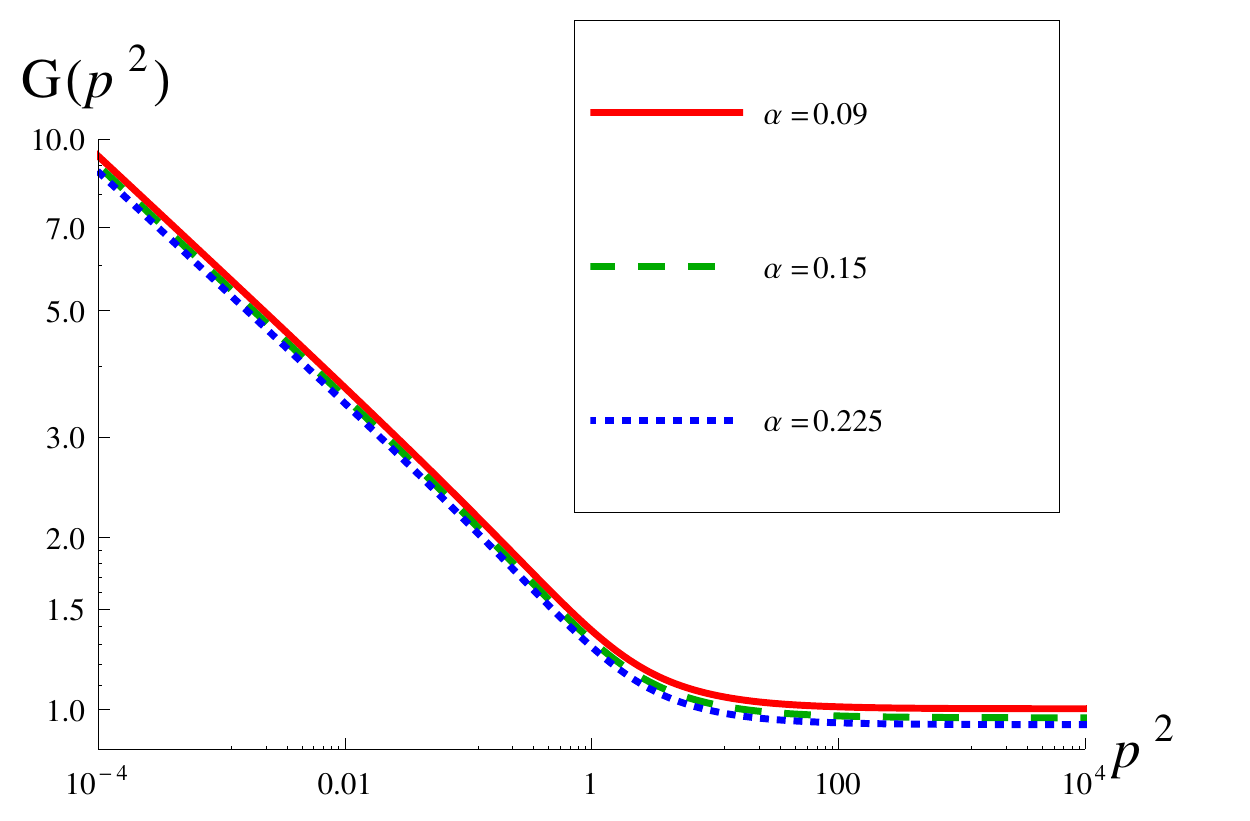}
 \end{minipage}
 \hfill
 \begin{minipage}[t]{0.48\textwidth}
 \includegraphics[width=\textwidth]{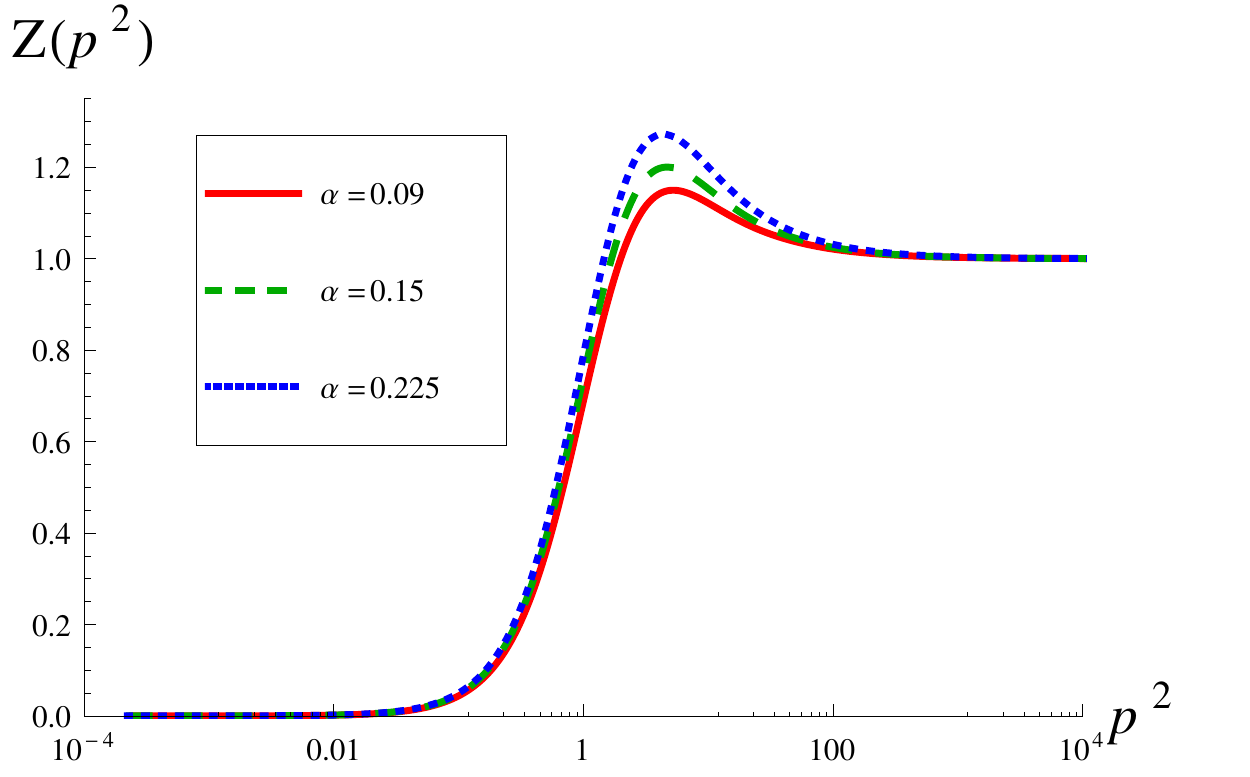}
 \end{minipage}
 \caption{\label{fig:props_bareGhg}Propagator dressing functions using a bare ghost-gluon vertex and the three-gluon vertex ansatz given in \eref{eq:3g-ans-3d} with $\alpha=0.09, 0.15, 0.225$. \textit{Left:} Ghost dressing function. In the UV the values are $0.94$, $0.97$ and $1$, respectively. \textit{Right:} Gluon dressing function.}
 \end{center}
\end{figure}

The use of a bare ghost-gluon vertex is standard in three and four dimensions, see, for example, \cite{Zwanziger:2001kw,Lerche:2002ep,Fischer:2003rp,Maas:2004se,Boucaud:2008ji,Aguilar:2008xm,Aguilar:2010zx}. The small deviations from the tree-level observed by lattice calculations \cite{Cucchieri:2004sq,Cucchieri:2008qm,Cucchieri:2006tf,Ilgenfritz:2006he} do not affect the IR or UV behavior and influence only the mid-momentum regime. Various continuum analyses of the vertex \cite{Schleifenbaum:2004id,Boucaud:2011eh,Fister:2011uw} yielded qualitatively similar results as from lattice calculations. Therefore we start our investigations with a bare ghost-gluon vertex. For the three-gluon vertex we use the model of \eref{eq:3g-ans-3d} in analogy with three dimensions. It turns out that the ghost dressing function does not automatically approach one in the UV. However, since the dimension of the coupling constant is that of momentum we know from dimensional arguments that it should behave for large momenta as
\begin{align}
 G(p^2)\xrightarrow{p^2\rightarrow \infty} \frac{1}{1+c\, g^2/p^2}
\end{align}
with $c$ a constant. Also lattice calculations \cite{Maas:2007uv} confirm that the ghost should be close to one in the UV. In order for our solution to satisfy this criterion we adjust the parameter $\alpha$ of the three-gluon vertex model \eref{eq:3g-ans-3d}. The resulting propagator dressings are shown in \fref{fig:props_bareGhg} for several values of $\alpha$. Indeed we can find a value where the ghost dressing has the correct UV value, namely $\alpha=0.09$.

The reason for the sensitivity of the ghost propagator in the mid-momentum regime is a rather intricate relation in the ghost DSE. Its unsubtracted form has the structure
\begin{align}
 \frac{1}{G(p^2)}=1+\Sigma(p^2)\nonumber.
\end{align}
For large momenta, asymptotic freedom requires that $\Sigma(p^2)$ vanishes, and thus the ghost dressing function approaches necessarily one. At low momenta, $\Sigma(p^2)$ will behave in the scaling case as $a+A(p^2)^\ka+{\cal O}((p^2)^\delta)$ with some constant $a$ and $\delta>\ka$. Only when $a=-1$, the scaling solution can be a solution to the equation. This requires that $\Sigma$ must provide a large enough integrated strength. Due to the integral measure, the far infrared does not contribute to this, and due to asymptotic freedom neither does the ultraviolet. Thus, $a=-1$ must be provided by the mid-momentum behavior. This can be generated by two ingredients. One is the mid-momentum enhancement of the gluon dressing, its maximum. The other is due to an enhancement of the ghost-gluon vertex. 
In four dimensions this problem is absent because there it is fixed by wave-function renormalization \cite{Fischer:2008uz}.

\subsection{Ghost-gluon vertex model}

Since the ghost UV behavior is so sensitive to the mid-momentum interval of the integrals, we will now investigate the impact of improved vertex models as motivated by lattice data.
In the following we will only use the transversely projected gluon DSE, so we only need to model the dressing function $D^{A\bar{c}c}_t$ of the ghost-gluon vertex which we take as
\begin{align}\label{eq:ghg_ansatz}
 D^{A\bar{c}c}_{t,mod}(r,p,q)=1+\frac{1}{\Lambda^2+p^2+q^2+r^2}\left(f_{IR}+f_{IM}\frac{\Lambda^2(p^2+q^2)}{\Lambda^4+p^4+q^4}\right).
\end{align}
The new term proportional to $f_{IM}$ adds a bump in the mid-momentum regime as seen in lattice calculations \cite{Maas:2007uv,Cucchieri:2011ig}. This bump occurs in three and four dimensions likewise, see refs. \cite{Cucchieri:2004sq,Cucchieri:2008qm,Cucchieri:2006tf}. It was also reproduced by a semi-perturbative DSE calculation \cite{Schleifenbaum:2004id} and was observed in similar form at finite temperature \cite{Fister:2011uw}. The parameter $f_{IR}$ sets the IR value of the dressing and $\Lambda$ is a scale parameter. This form of the vertex leaves the value of $\ka$ unchanged as discussed in Sec.~\ref{sec:other_kappa}, but not the IR value of the coupling, which scales with $1/(1+f_{IR}/\Lambda^2)$.

\begin{figure}[tb]
 \begin{center}
 \begin{minipage}[t]{0.48\textwidth}
 \includegraphics[width=\textwidth]{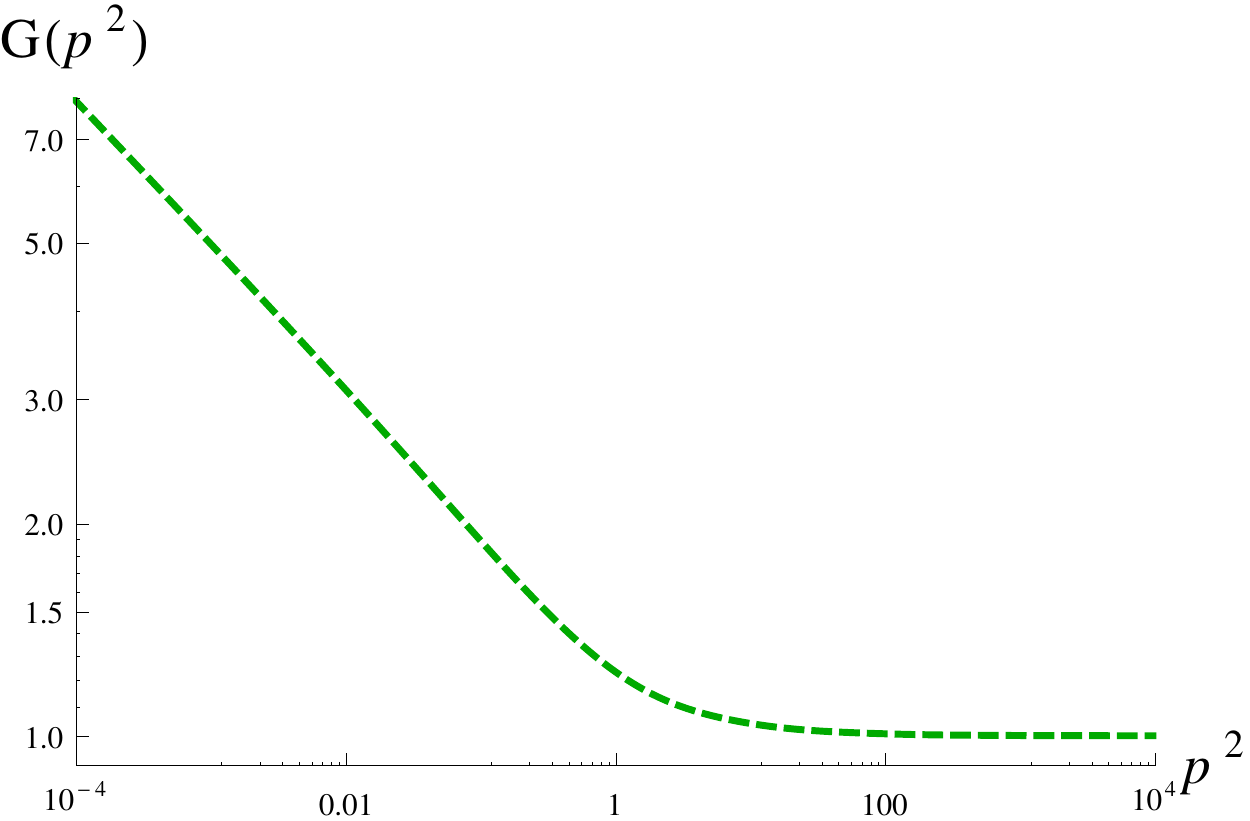}
 \end{minipage}
 \hfill
 \begin{minipage}[t]{0.48\textwidth}
 \includegraphics[width=\textwidth]{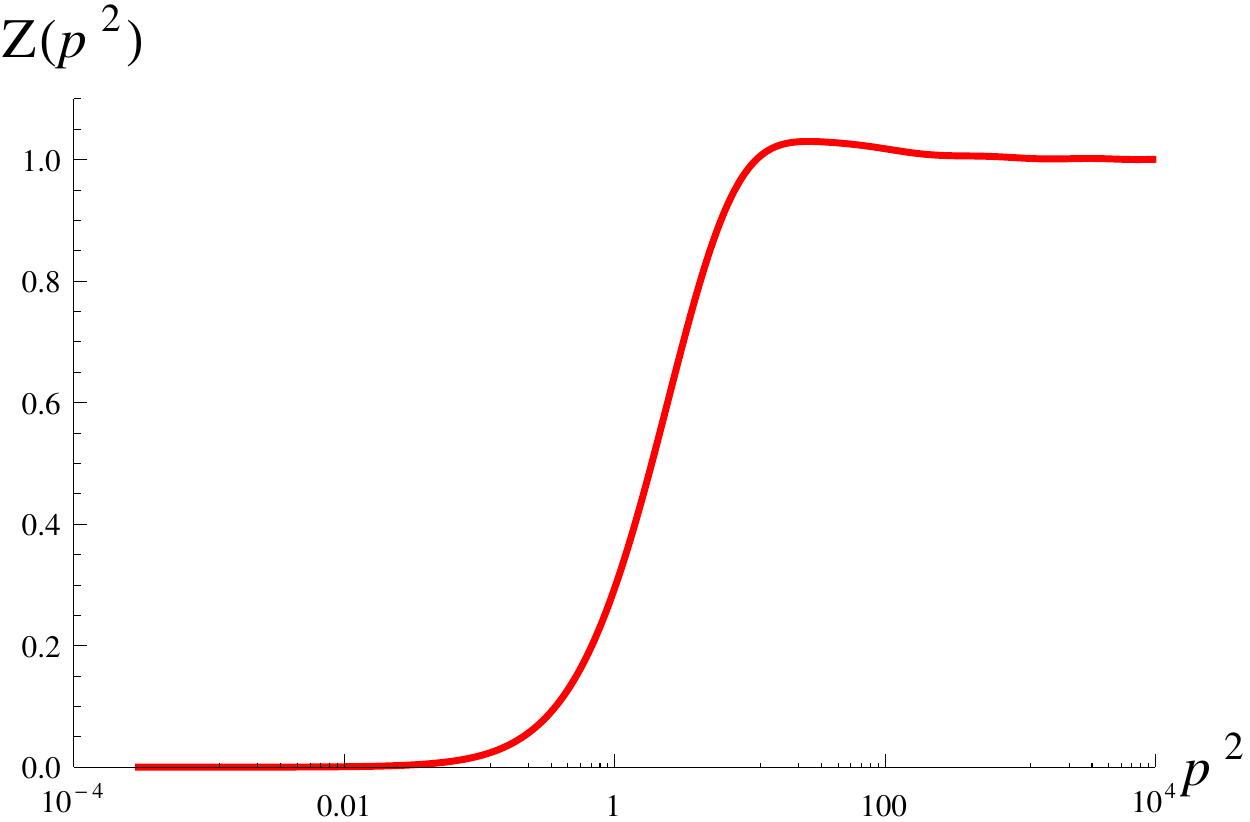}
 \end{minipage}
\caption{\label{fig:props_VModels}Ghost (\textit{left}) and gluon (\textit{right}) dressings obtained with the models given in eqs.~(\ref{eq:ghg_ansatz}) and (\ref{eq:tg_ansatz}) using the values of Table~\ref{tab:vertex_params} for the parameters.}
 \end{center}
\end{figure}

For the three-gluon vertex we will consider the form given by \eref{eq:tg_ansatz}, which is the lattice motivated model. Note that we use only one dressing function, but a posteriori our calculations show that this can describe lattice data rather well, see Appendix~\ref{sec:tgvert}.
Fixing the parameters of the ghost-gluon vertex it is indeed possible to tune the three-gluon vertex parameters such that the ghost dressing function becomes one in the UV. A possible choice of parameter values is given in Table~\ref{tab:vertex_params} and the resulting dressing functions are plotted in \fref{fig:props_VModels}.

\begin{table}[tb]
\begin{minipage}[t]{0.48\textwidth}
\begin{center} 
\begin{tabular}{|l|l|}
\hline
Parameter & Value\\
\hline\hline
$\Lambda$ & $1$\\
\hline
$f_{IR}$ & $2.14189$\\
\hline
$f_{IM}$ & $0.6$\\
\hline
$h_{IR} $ & $-109.616/-54.2$\\
\hline
$h_{IM}$ & $9.88$\\
\hline
\end{tabular}
\caption{\label{tab:vertex_params}Values employed for the parameters of the ghost-gluon and three-gluon vertex ans\"atze given in eqs.~(\ref{eq:ghg_ansatz}) and (\ref{eq:tg_ansatz}), respectively. The second value for $h_{IR}$ is for $SU(2)$.}
\end{center}
\end{minipage}
\hfill
\begin{minipage}[t]{0.48\textwidth}
\begin{center} 
\begin{tabular}{|l|l|}
\hline
Parameter & Value\\
\hline\hline
$\Lambda$ & $1$\\
\hline
$h_{IR} $ & $-31.5/-14.75$\\
\hline
$h_{IM}$ & $9.88$\\
\hline
\end{tabular}
\caption{\label{tab:3g_params_dynGhg}Values employed for the parameters of the three-gluon vertex ansatz given in \eref{eq:tg_ansatz} when including the ghost-gluon vertex dynamically. The second value for $h_{IR}$ is for $SU(2)$.}
\end{center} 
\end{minipage}
\end{table}

A remarkable feature of these and the following results is that they cure the infrared singularities of perturbation theory  due to the infrared suppression of the gluon propagator. For the ghost-loop in the gluon equation, however, this is naively unexpected, given that the ghost propagator is infrared more divergent than in perturbation theory. The explanation is that intricate cancellations in the angular loop integral occur which are due to the non-trivial momentum dependences of the dressing functions, emphasizing the delicate balancing in scaling type solutions.

\section{Including the ghost-gluon vertex dynamically}
\label{sec:ghgvert}

Knowing how important the three-point functions are for the correct UV behavior of the ghost propagator we will now include the ghost-gluon vertex dynamically into our calculations. Working in two dimensions is thereby advantageous because we only have to calculate two-dimensional integrals, whereas in three and four dimensions the integrals are three-dimensional. Furthermore, in contrast to four dimensions, the UV behavior is trivial. This calculation is therefore also an exploratory study for future calculations in four dimensions to extend the currently employed truncation schemes beyond the propagator level.

We will again use the transversely projected gluon DSE so that only
one dressing function of the ghost-gluon vertex is
relevant. Three-point functions depend on three variables for which we
choose the squares of the two ghost momenta and the angle $\varphi$ between them. The vertex is calculated for a grid in these variables. For intermediate points we use linear interpolation. If any momentum is outside the grid we use the value of the dressing at its boundary. Considering the increased complexity of the ghost-gluon vertex DSE it was advantageous to derive the kernels with the program \textit{DoFun} \cite{Alkofer:2008nt,Huber:2011qr}. For solving the DSEs the framework provided by \textit{CrasyDSE} \cite{Huber:2011xc} was used.

\begin{figure}[tb]
 \begin{center}
 \includegraphics[width=\textwidth]{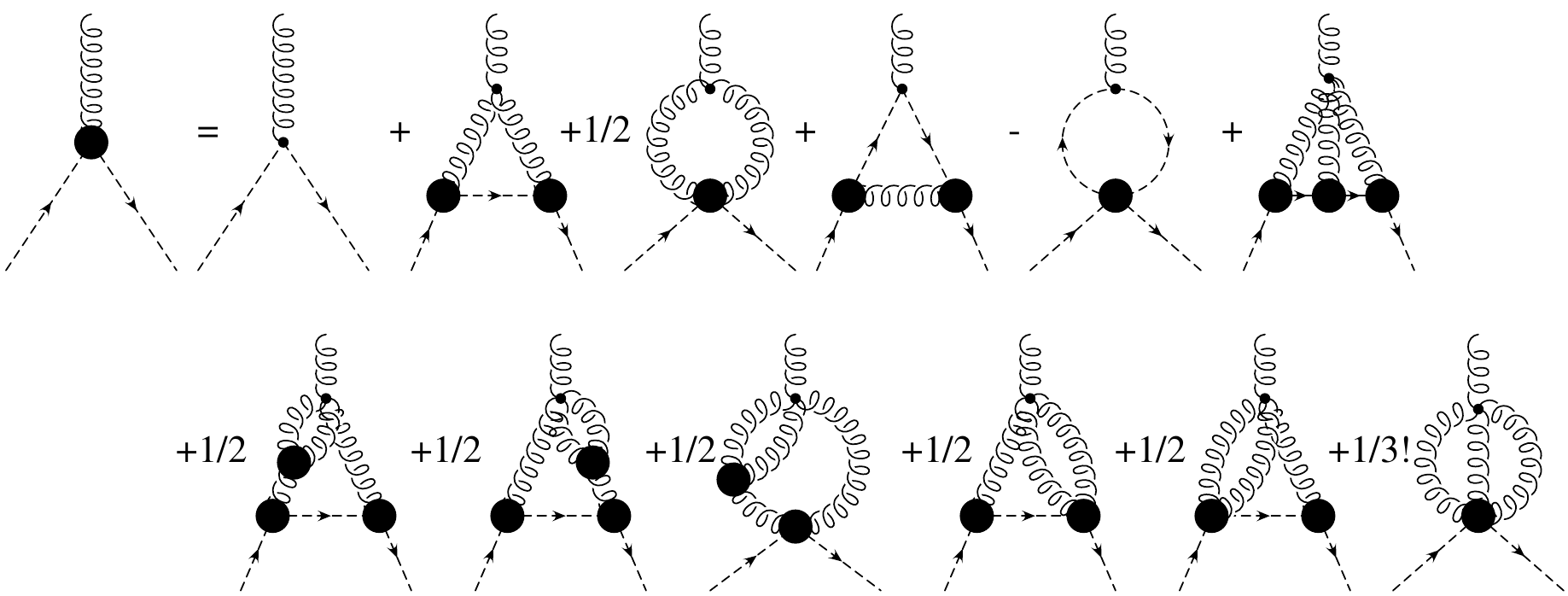}
\caption{\label{fig:ghg-DSE}The ghost-gluon vertex DSE. All internal propagators are dressed. Thick blobs denote dressed vertices. Wiggly lines are gluons, dashed ones ghosts. The employed truncation consists of the first, the second and the fourth diagrams.}
 \end{center}
\end{figure}

The ghost-gluon vertex has two distinct DSEs, which differ by the field that is attached to the bare vertex. We take the DSE where this is the gluon field. Although this DSE seems more complicated than the other one, since it has more terms, it turns out that in our truncation scheme it is simpler. The full DSE is shown in \fref{fig:ghg-DSE}. 
Our truncation of the DSE is motivated again by keeping the leading IR and UV contributions. The former can be identified by power counting \cite{Huber:2007kc} and the latter by counting the powers of the coupling.
Because of the UV argument we can discard all diagrams containing a bare four-gluon vertex, i.~e., all two-loop diagrams, and also the third and fifth diagrams on the right-hand side. Note that the last one formally occurs at leading order in the IR, but it does not represent one of the main contributions there, because when we insert the DSE of the irreducible quartic ghost vertex, we see that the IR leading contribution is a two-loop diagram which should give only a minor correction to the IR behavior of the ghost-gluon vertex. As verified by our calculations below already the fourth diagram only yields small corrections to the bare vertex.

\begin{figure}[tb]
 \begin{center}
 \begin{minipage}[t]{0.48\textwidth}
 \includegraphics[width=\textwidth]{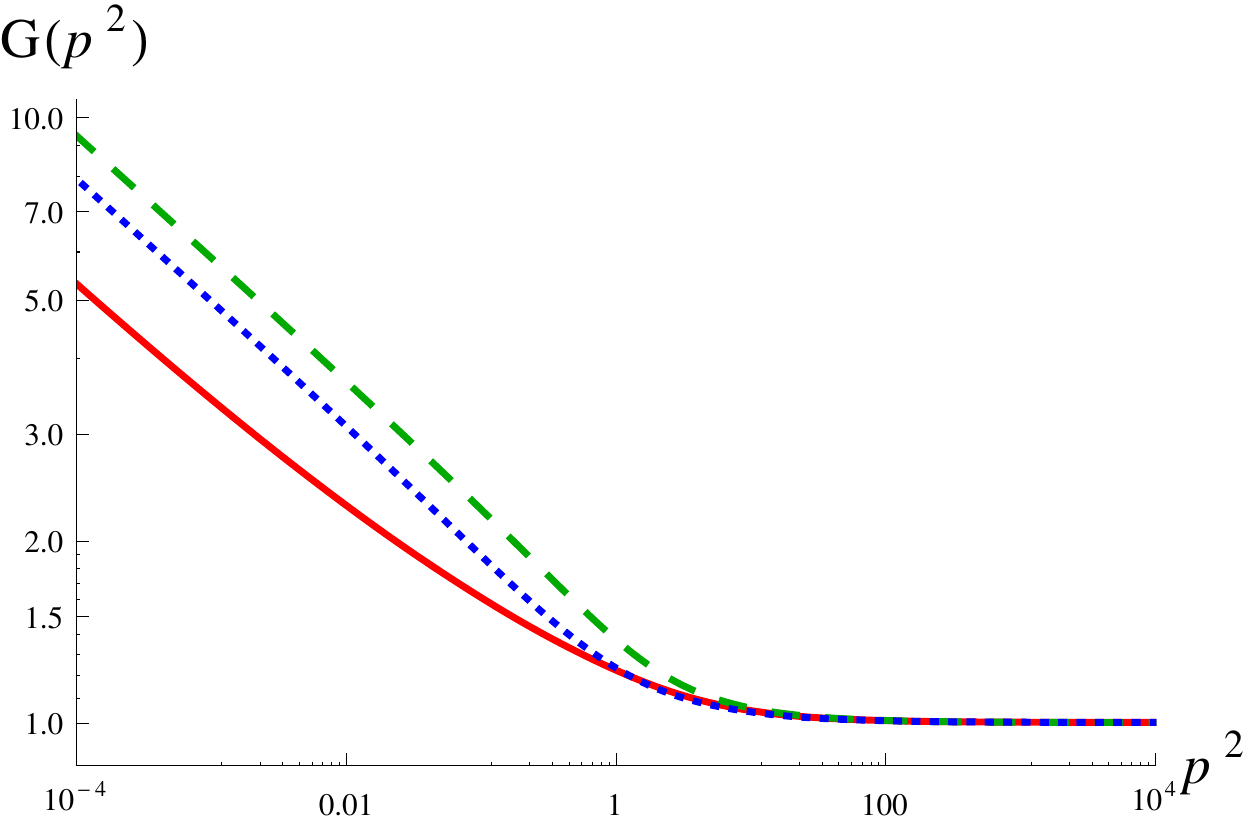}\\
 \includegraphics[width=\textwidth]{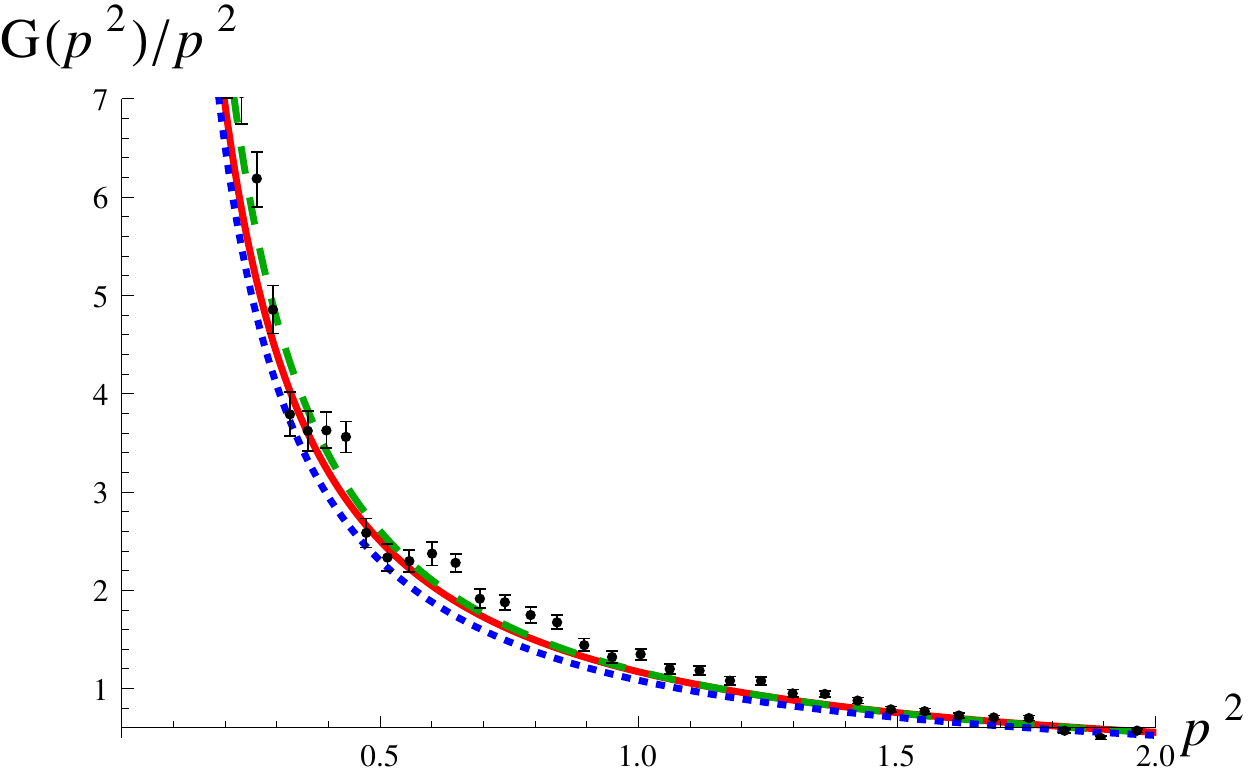}
 \end{minipage}
 \hfill
 \begin{minipage}[t]{0.48\textwidth}
 \includegraphics[width=\textwidth]{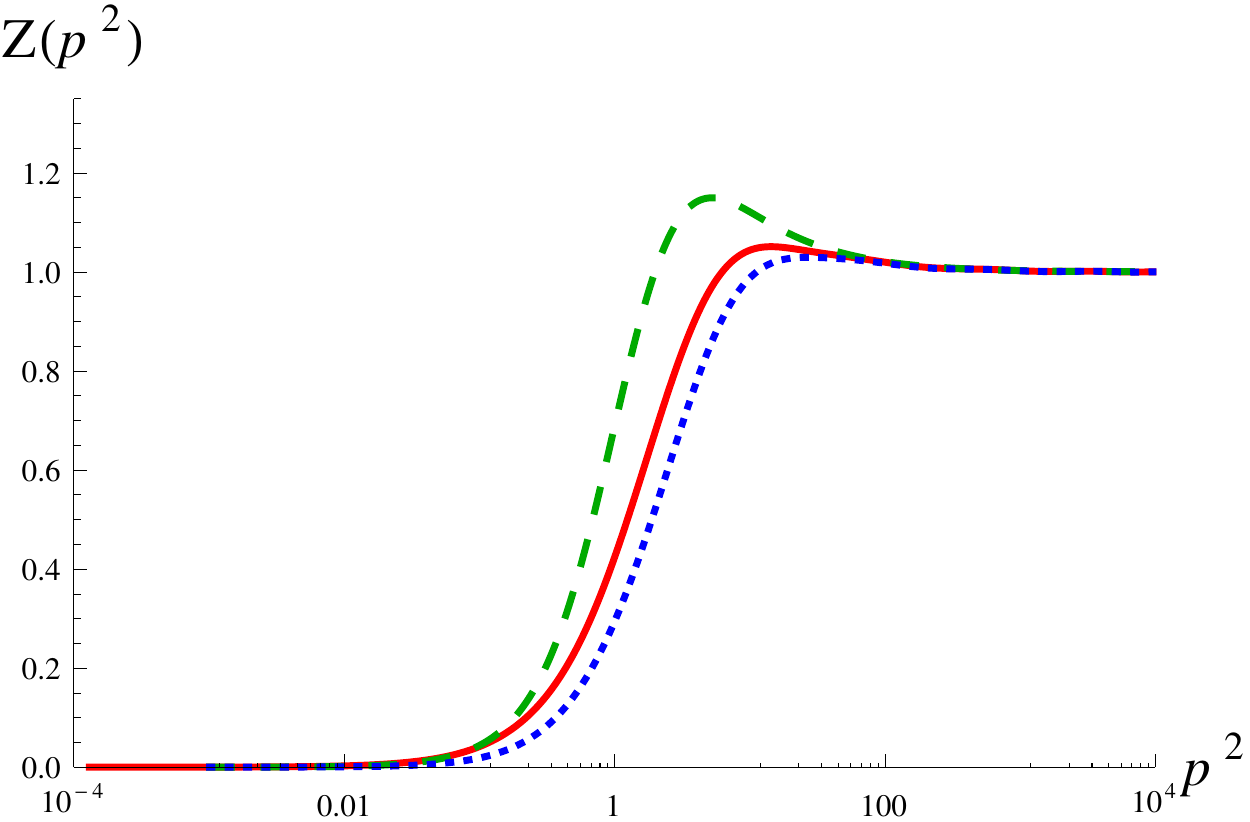}
\includegraphics[width=\textwidth]{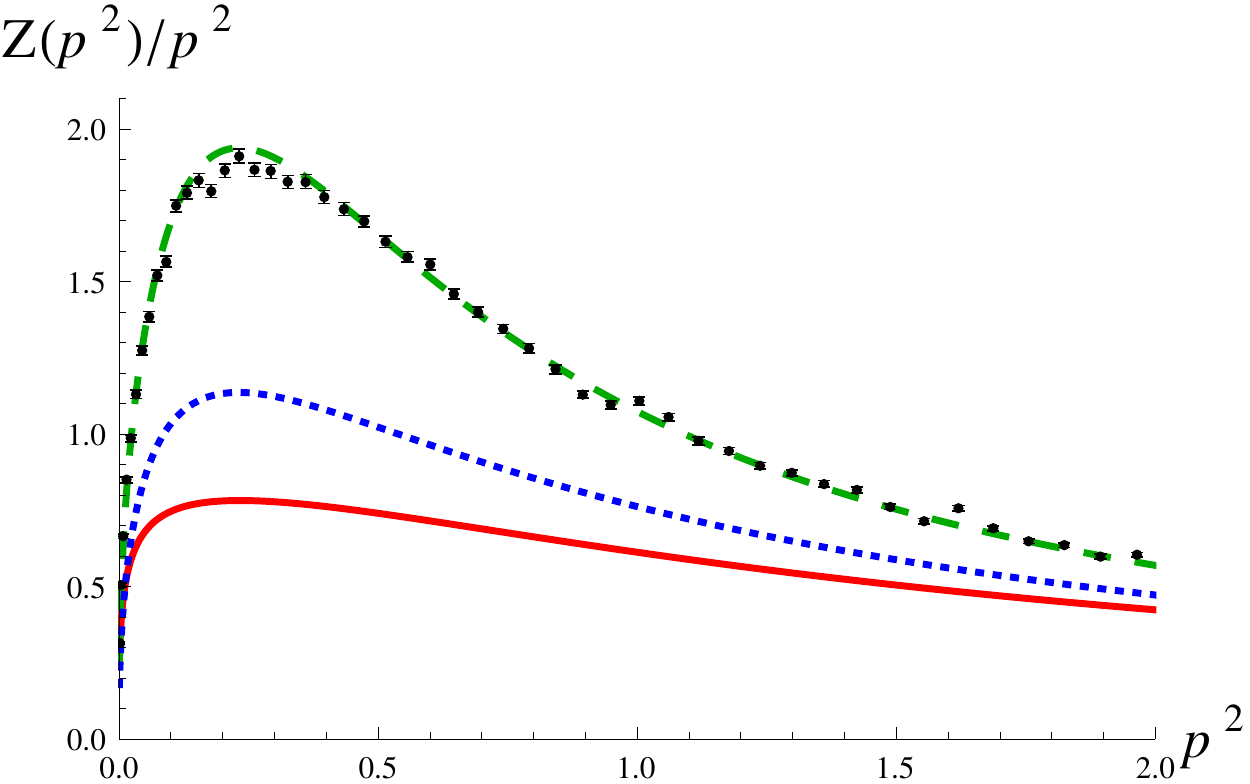}
 \end{minipage}
\caption{\label{fig:props_DynGhg}Ghost (\textit{left}) and gluon
  (\textit{right}) dressings obtained when including the ghost-gluon
  vertex dynamically (red/solid lines) and using the three-gluon
  vertex ansatz from \eref{eq:tg_ansatz} with parameters given in
  Table~\ref{tab:3g_params_dynGhg} compared to the results when using
  a bare ghost-gluon vertex (green/dashed lines) or the model from
  \eref{eq:ghg_ansatz} (blue/short-dashed line) with parameters given
  in Table~\ref{tab:vertex_params}. \textit{Bottom:} Comparison of the
  propagators for $SU(2)$ with lattice data (black points) from \cite{Maas:2011se}. In the plot of the ghost propagators
  the continuum results are nearly indistinguishable. Scale set by
  matching the positions of the maxima in the gluon propagators. Note
  that this rescaling is responsible for the interchange of the
  blue/short-dashed line with the red/solid line.} 
 \end{center}
\end{figure}

\begin{figure}[tb]
 \begin{center}
 \begin{minipage}[t]{0.47\textwidth}
 \includegraphics[width=\textwidth]{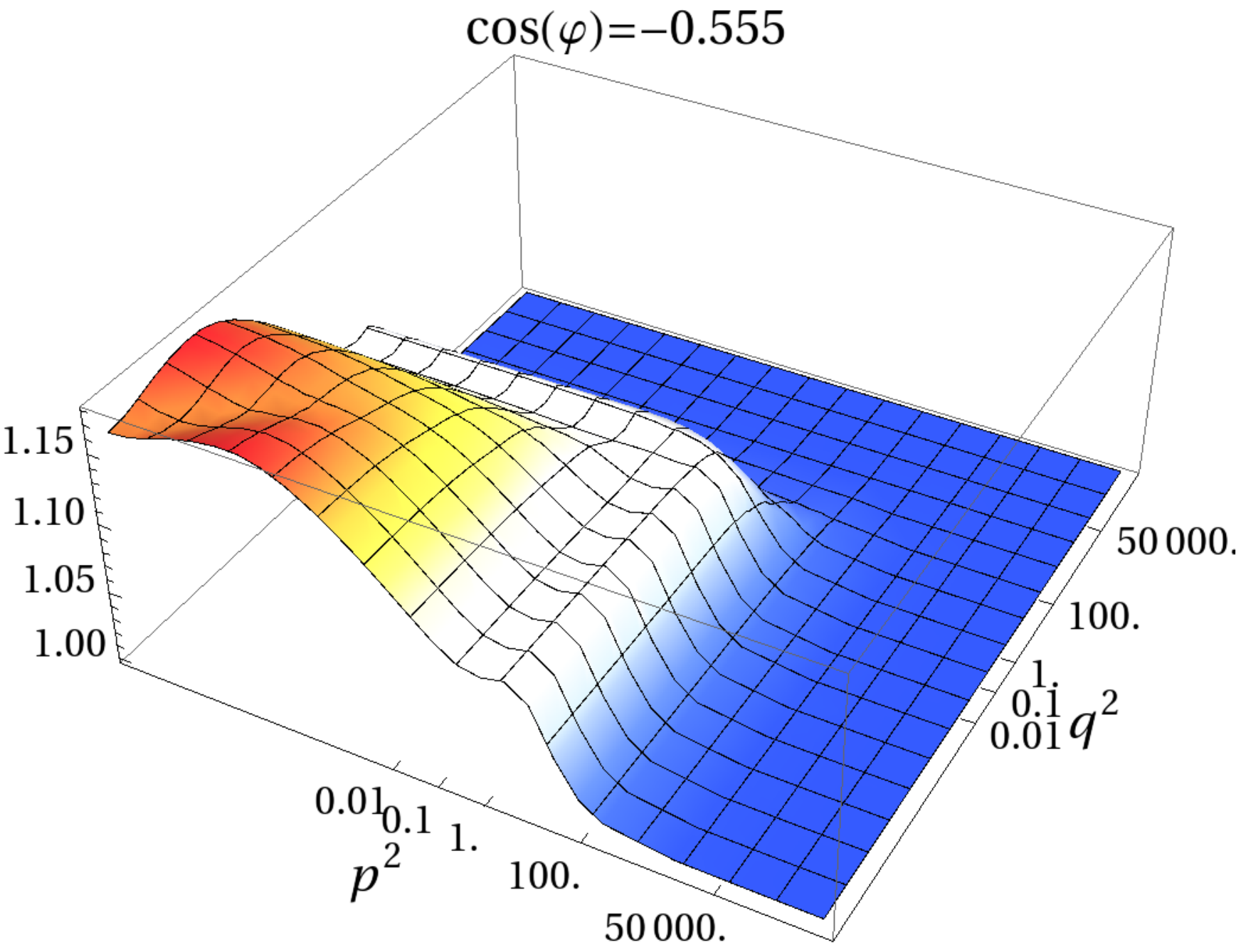}
 \includegraphics[width=\textwidth]{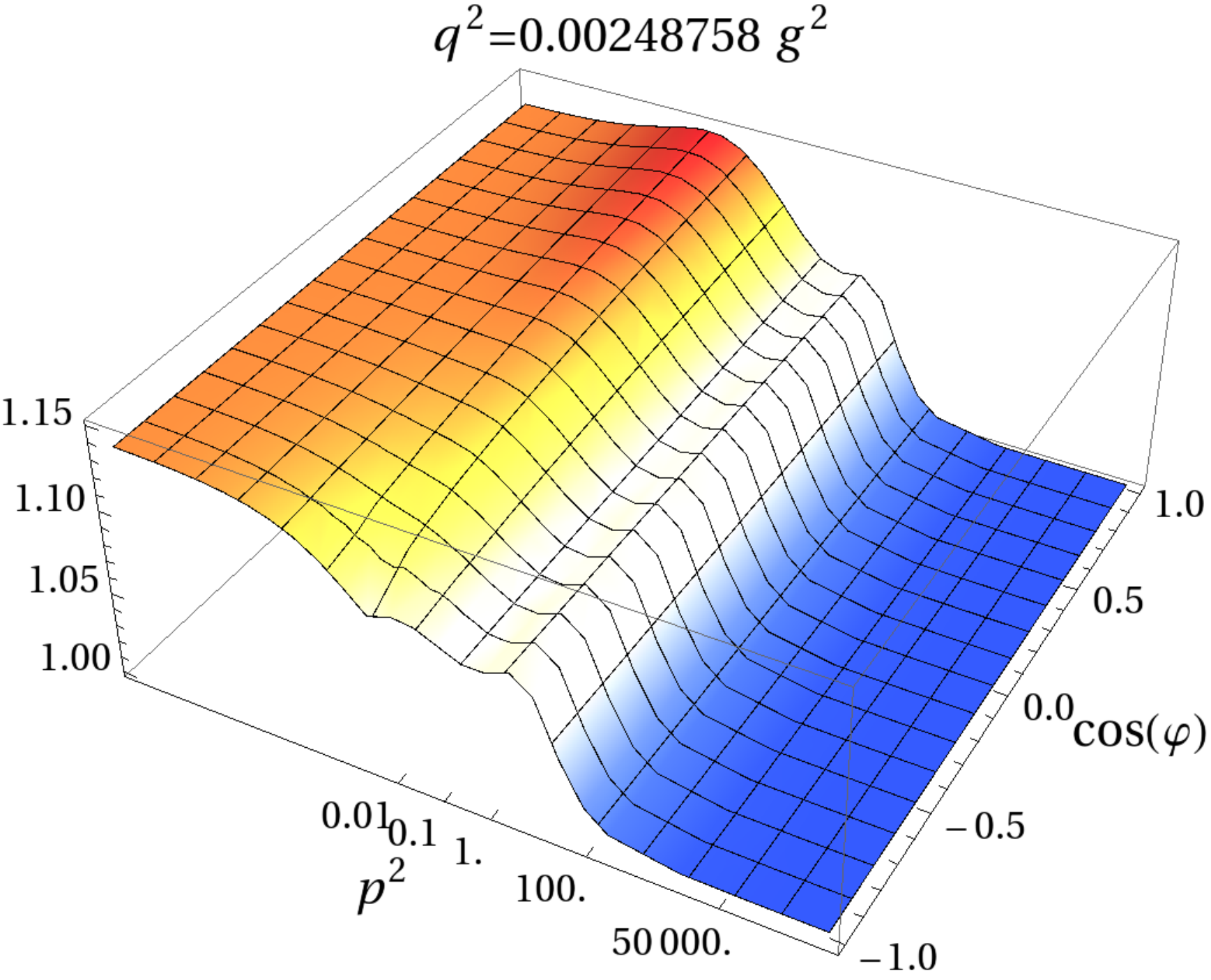}
 \includegraphics[width=\textwidth]{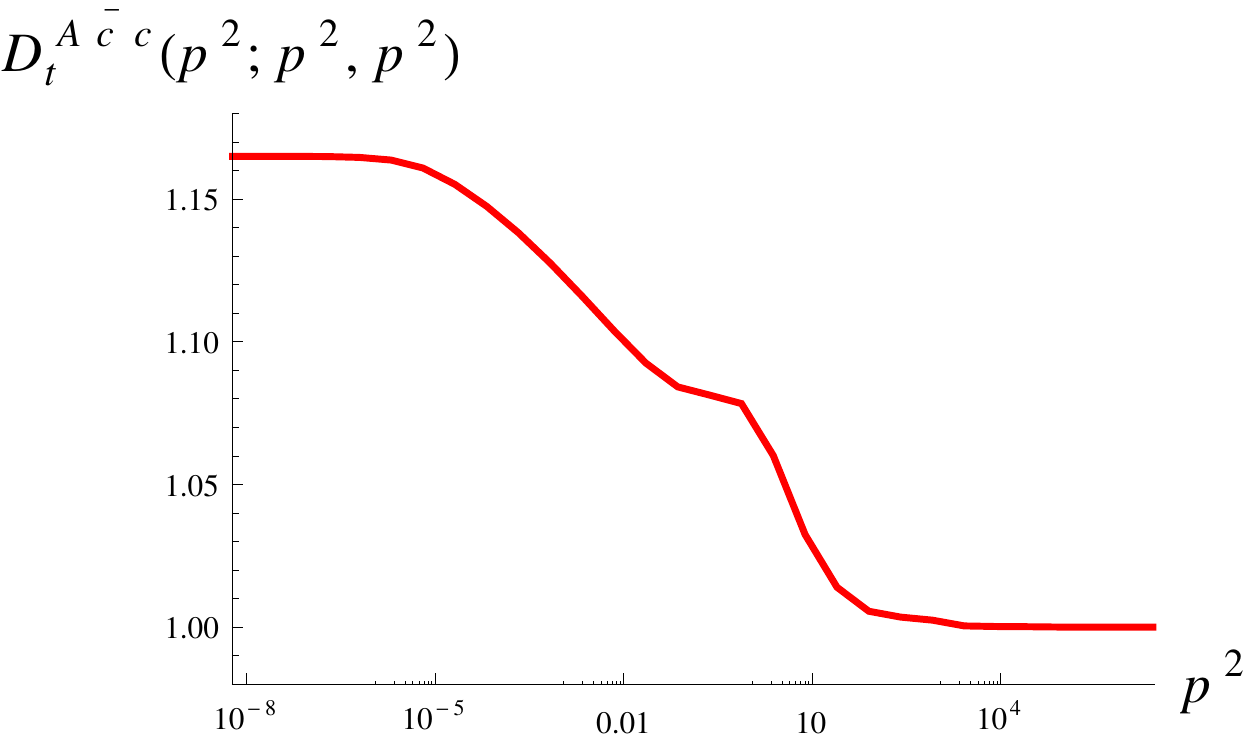}
 \end{minipage}
 \hfill
 \begin{minipage}[t]{0.47\textwidth}
 \includegraphics[width=\textwidth]{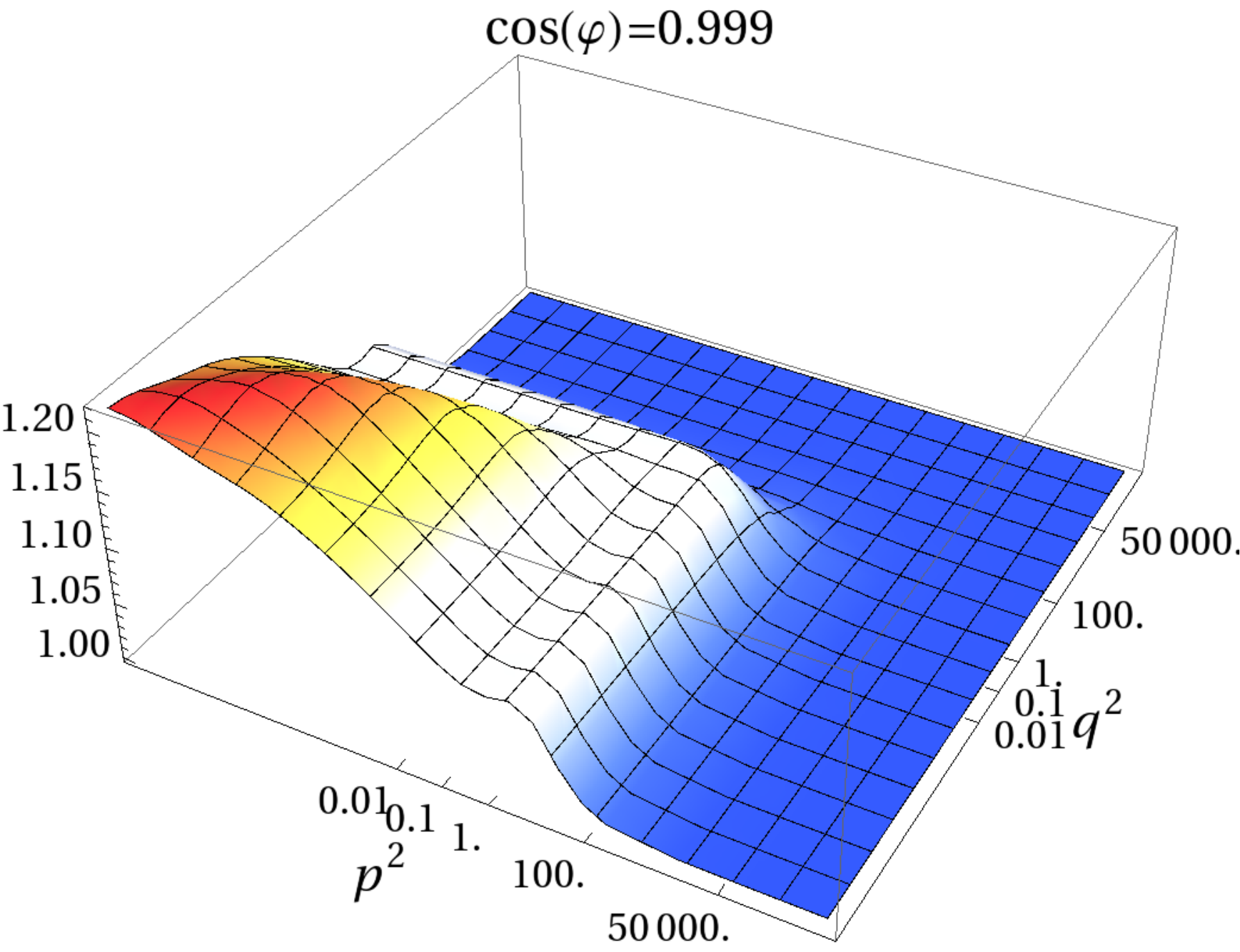}
 \includegraphics[width=\textwidth]{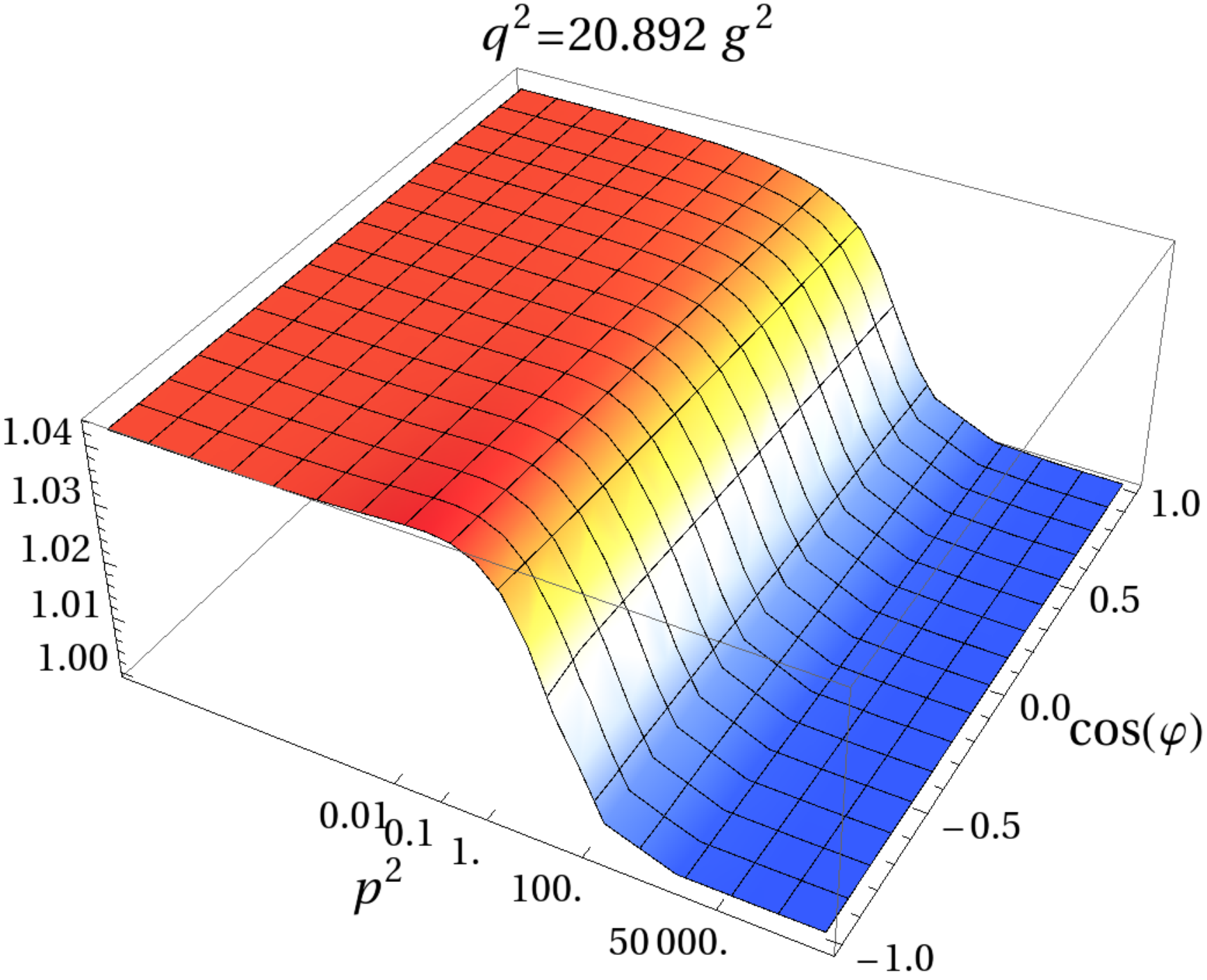}
 \includegraphics[width=\textwidth]{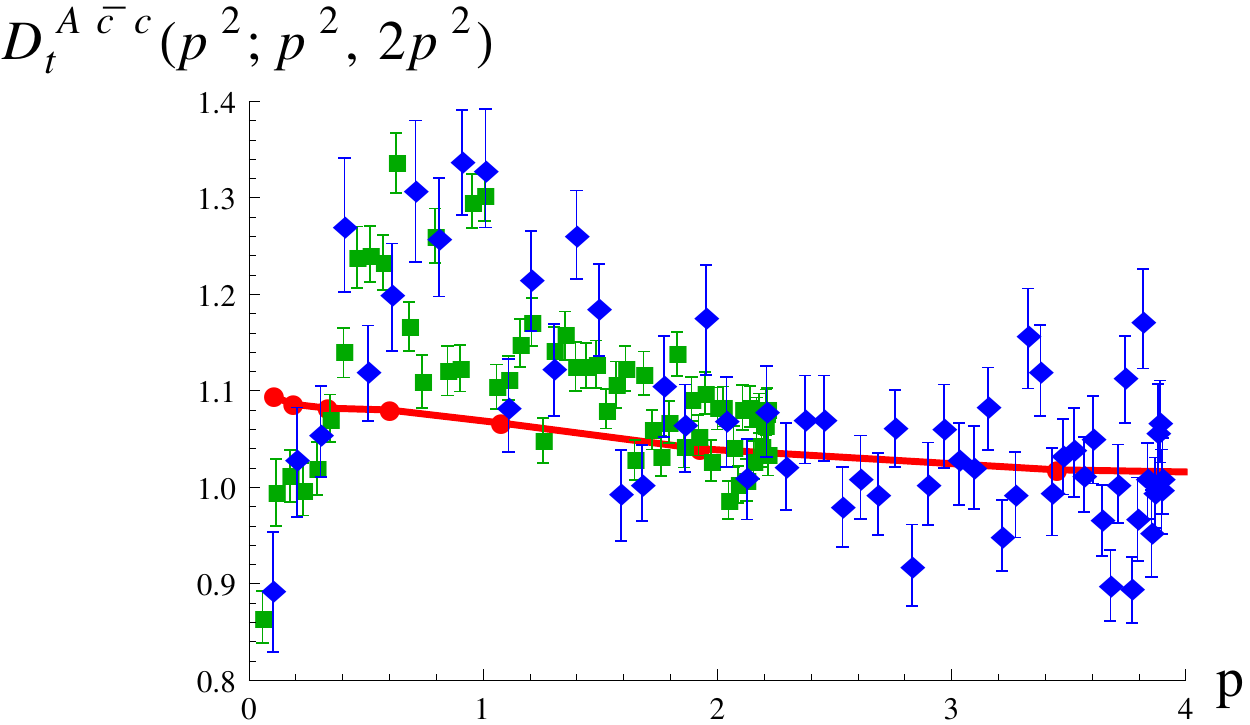}
 \end{minipage}
\caption{\label{fig:ghg_DynGhg}Dressing of the ghost-gluon vertex for various momentum configurations obtained from the coupled system of propagators and the vertex itself. For the three-gluon vertex the ansatz of \eref{eq:tg_ansatz} with the parameters of Table~\ref{tab:3g_params_dynGhg} was used. \textit{Top:} Fixed angle as indicated at the top of each plot. \textit{Middle:} Fixed momentum $q^2$ as indicated at the top of each plot. Note the different scales on the $z$-axes. \textit{Bottom:} On the left the symmetric configuration is shown. Since this configuration cannot be realized on the lattice, no comparison with lattice data is possible. The bump stems from the non-Abelian diagram. On the right the gluon momentum and one ghost momentum are orthogonal. The red/solid line is the result from the DSE calculation and green squares are for $\beta=10$/$L=21fm^{-1}$ and blue diamonds for $\beta=22.5$/$L=12fm^{-1}$ lattice data from ref. \cite{Maas:2007uv}.}
 \end{center}
\end{figure}

The resulting dressings of the propagators and the ghost-gluon vertex are shown in Figs.~\ref{fig:props_DynGhg} and \ref{fig:ghg_DynGhg}, respectively. For comparison the plots also contain results presented in the previous sections. The differences originating in the mid-momentum regime can clearly be seen. 
Most notably the maximum of the gluon dressing function is driven to even larger momenta and becomes more shallow. In the ghost-gluon vertex dressing also a bump in the mid-momentum regime is seen. However, it can be uniquely traced back to the second diagram on the right-hand of its DSE in \fref{fig:ghg-DSE}. We also tried the second version of the ghost-gluon vertex DSE, but we did not obtain a solution. The reason is that in this DSE the dressed instead of the bare three-gluon vertex appears and introduces an instability in the system of equations. This again illustrates the delicate balancing in the mid-momentum regime in two dimensions.

In \fref{fig:props_DynGhg} we also compare the results to lattice data. For this we redid the calculations for $SU(2)$, since the lattice data was obtained for two colors. Most notably the results from our first set-up with a bare ghost-gluon vertex lie almost on top of the lattice points. However, this can be considered merely a lucky coincidence since we have explicitly demonstrated the sensitivity to the mid-momentum regime before, and the employed vertex ans\"atze in this set-up do not mimic the correct behavior in all momentum regions. Improving the vertex models has the consequence that in the mid-momentum regime a gap between the DSE and lattice results opens as it is also known from four dimensions. The dynamical inclusion of the ghost-gluon vertex increases this effect even more. This is not unexpected since we discarded all two-loop diagrams. Unfortunately the inclusion of two-loop diagrams requires an extension of currently employed methods and their implementation in DSEs has not been explored thoroughly enough to add them here straightforwardly, but see, for example, \cite{Bloch:2003yu,Alkofer:2011di}.
Thus, although in two dimensions the qualitative properties of the solutions of the DSEs are simpler and less ambiguous, the quantitative features are not described as satisfactorily as in three and four dimensions. In four dimensions this difference is likely due to the possibility of renormalization, which permits to shift various effects to different momentum scales. In three dimensions, where one also finds quantitatively acceptable descriptions of the mid-momentum regime, it is probably due to the increased contribution from the momentum integral measure, which permits rather small differences at larger momenta to have already a significant impact.

\section{Conclusions}
\label{sec:conclusions}

Summarizing, we have provided the first full solution of the
two-dimensional DSEs of Landau gauge Yang-Mills theory, including the equation for the propagators {\em and} the ghost-gluon vertex. The inclusion of the latter extends currently employed truncation schemes from four dimensions and was thus expected to reduce the truncation dependence. However, we found that the mixing of different momentum regimes in superrenormalizable theories invalidates the simple truncations used in the four-dimensional, renormalizable case. The reasons for this can be understood from the intricate cancellations necessary for a scaling type solution, which we find, in accordance with \cite{Cucchieri:2012cb}, to be the only viable type of solution in two dimensions. That a similar situation did not arise in this severeness in earlier studies in three dimensions \cite{Maas:2004se} must be attributed with hindsight to quantitative effects.

As a consequence, the results here have not yet reached a maturity as in higher dimensions when it comes to quantitatively reproducing lattice results. This will require much more sophisticated truncation schemes, which will likely require the inclusion of two-loop terms. In DSEs, this is a formidable endeavor, and thus a renormalization group approach, with its intrinsic one-loop structure, may be an interesting alternative. Nonetheless, we reproduced all qualitative features and provided an understanding of the underlying mechanisms.

\section*{Acknowledgments}
M.Q.H.\ was supported by the Alexander von Humboldt
foundation, A.M.\ by the DFG under grant number MA
3935/5-1, and L.vS.\ by the Helmholtz International Center for FAIR
within the LOEWE program of the State of Hesse, the Helmholtz
Association Grant VH-NG-332, and the European Commission,
FP7-PEOPLE-2009-RG No. 249203. Plots of DSEs were created with
\textit{FeynDiagram} and \textit{JaxoDraw} \cite{Binosi:2003yf}.

\appendix

\section{Kernels}
\label{sec:app_kernels}

The kernel $K_{G}$ of the ghost DSE given in \eref{eq:gh-DSE} is
\begin{align}
 K_{G}(p,q)=\frac{ \left(x^2+(y-z)^2-2 x (y+z)\right)}{4 x y^2 z}
\end{align}
with $x=p^2$, $y=q^2$ and $z=(p+q)^2$.
The two kernels $K_{Z}^{gh}$ and $K_{Z}^{gl}$ of the gluon DSE \eref{eq:gl-DSE} read
\begin{align}
 K_{Z}^{gh}(p,q)&=-\frac{ x^2+(y-z)^2-2 x (y+z)}{4 x^2 y z},\nnnl
 K_{Z}^{gl}(p,q)&=\frac{x^4-8 x y z (y+z)+x^2 \left(-2 y^2-8 y z-2 z^2\right)+(y-z)^2 \left(y^2+2 y z+z^2\right)}{8 x^2 y^2 z^2}.
\end{align}
In order to get rid of the logarithmic divergences in the gluon DSE without using counter terms, the following expression is added to the kernel of the gluon loop $K_Z^{gl}$: 
\begin{align}
 K_Z^{gl,sub}(p,q)=\frac{1}{2 x y}.
\end{align}
The kernels for the ghost-gluon vertex are rather lengthy and not reproduced here. They were generated automatically using the programs \textit{DoFun} \cite{Alkofer:2008nt,Huber:2011qr} and {\it CrasyDSE} \cite{Huber:2011xc} using the ans\"atze for the vertices as described in the main text.

\section{Three-gluon vertex}
\label{sec:tgvert}

For solving the coupled system of propagators and ghost-gluon vertex in Sec.~\ref{sec:ghgvert} we employed a lattice inspired ansatz for the three-gluon vertex. Here we calculate the three-gluon vertex using the results obtained there for illustration and comparison. A combined calculation of propagators and both three-point functions did not yield a stable iteration which is again due to the effect truncations have on the mid-momentum regime. Nevertheless it is interesting to see that the lattice results can be reproduced at least qualitatively. Whereas the correct IR behavior should emerge automatically, the most important question is if the zero crossing as observed on the lattice is reproduced \cite{Maas:2007uv}.

\begin{figure}[tb]
 \begin{center}
  \includegraphics[width=\textwidth]{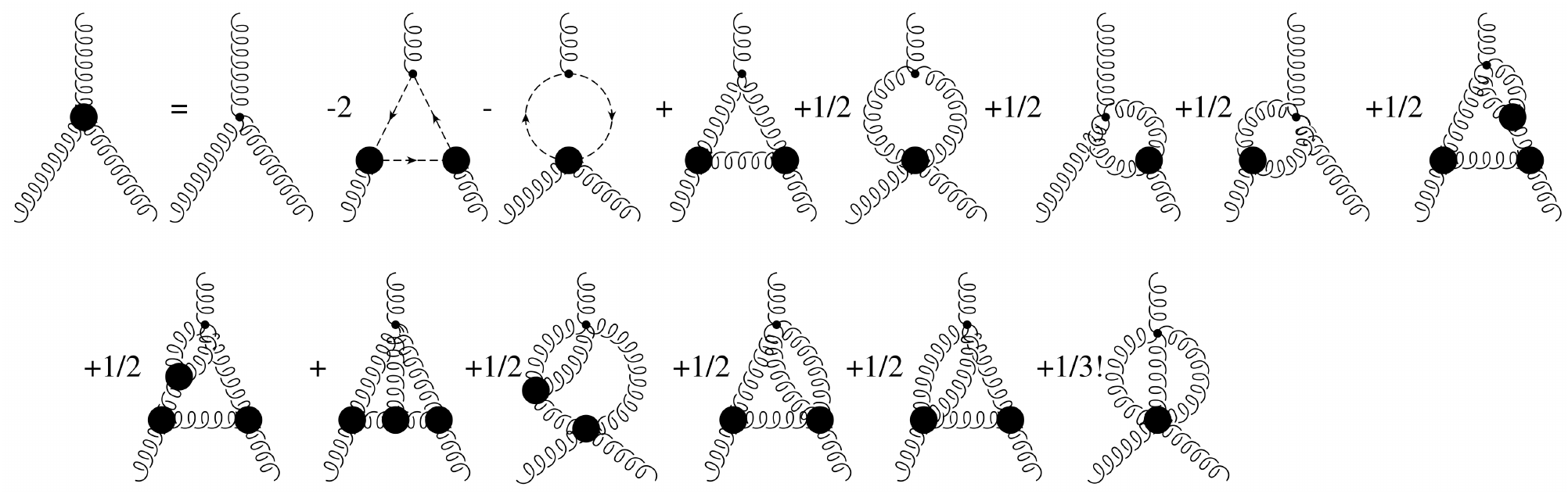}
  \caption{\label{fig:3g-DSE}The three-gluon vertex DSE. All internal propagators are dressed. Thick blobs denote dressed vertices. Wiggly lines are gluons, dashed ones ghosts.}
 \end{center}
\end{figure}

The DSE for the three-gluon vertex is depicted in \fref{fig:3g-DSE}.
In analogy to the truncation for the ghost-gluon vertex we discarded all diagrams with two loops or non-primitive vertices. This provides the correct IR behavior and includes also the leading corrections to the tree-level behavior in the UV. The remaining loops are the so-called ghost and gluon triangles represented by the second and fourth diagrams, respectively, on the right-hand side in \fref{fig:3g-DSE} and the swordfish diagrams represented by the fifth, sixth and seventh diagrams. For the dressed four-gluon vertex we use the bare one. This does not affect the IR and UV behavior but only the mid-momentum regime. Another approximation is that we use the result for the dressing of the three-gluon vertex as given in \eref{eq:tg_proj} below on the right-hand side of the three-gluon vertex DSE, i.~e., $\Gamma^{A^3}=D^{A^3}_{proj}$.

\begin{figure}[tb]
 \begin{center}
 \begin{minipage}[t]{0.48\textwidth}
 \includegraphics[width=\textwidth]{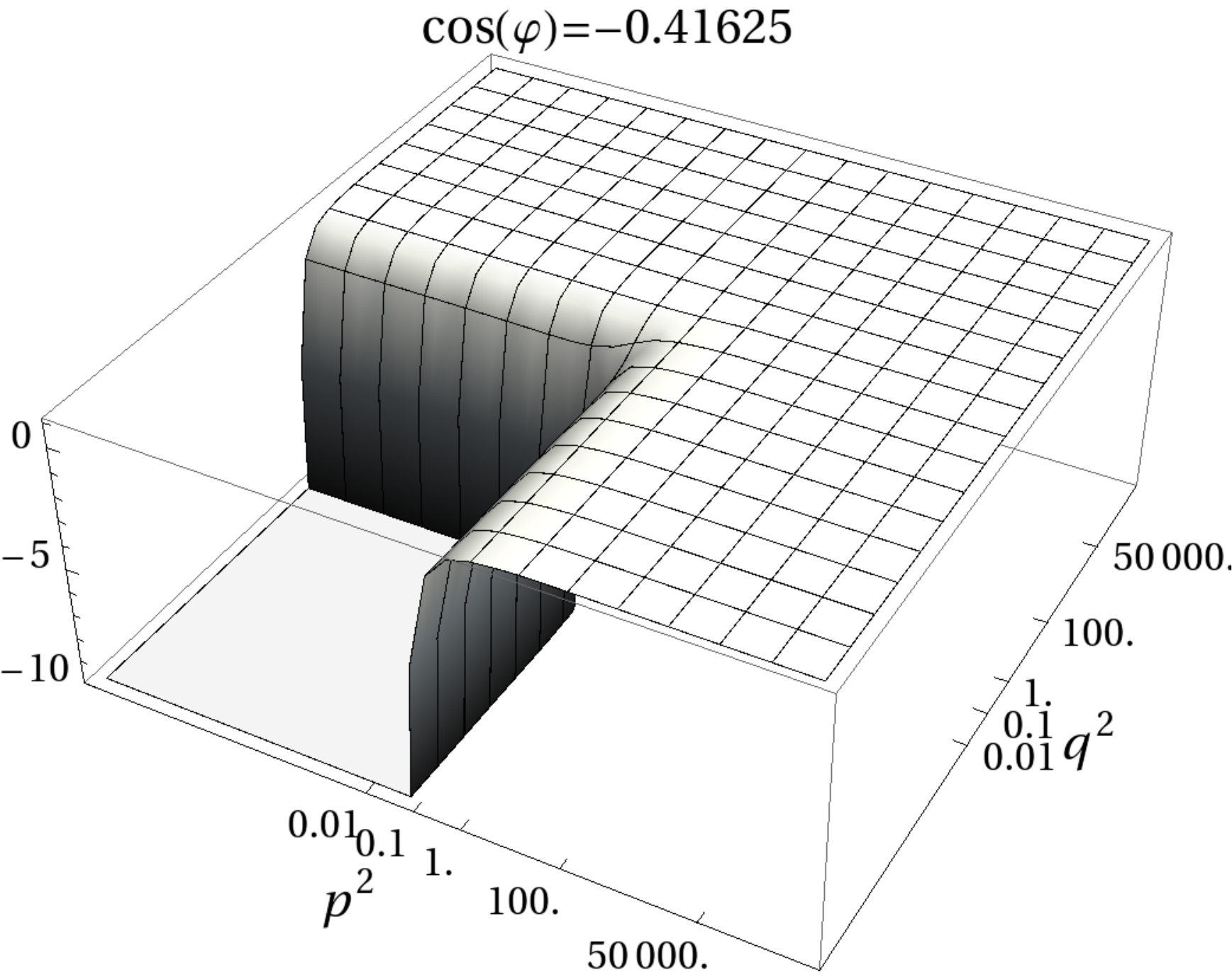}
 \end{minipage}
 \hfill
 \begin{minipage}[t]{0.48\textwidth}
 \includegraphics[width=\textwidth]{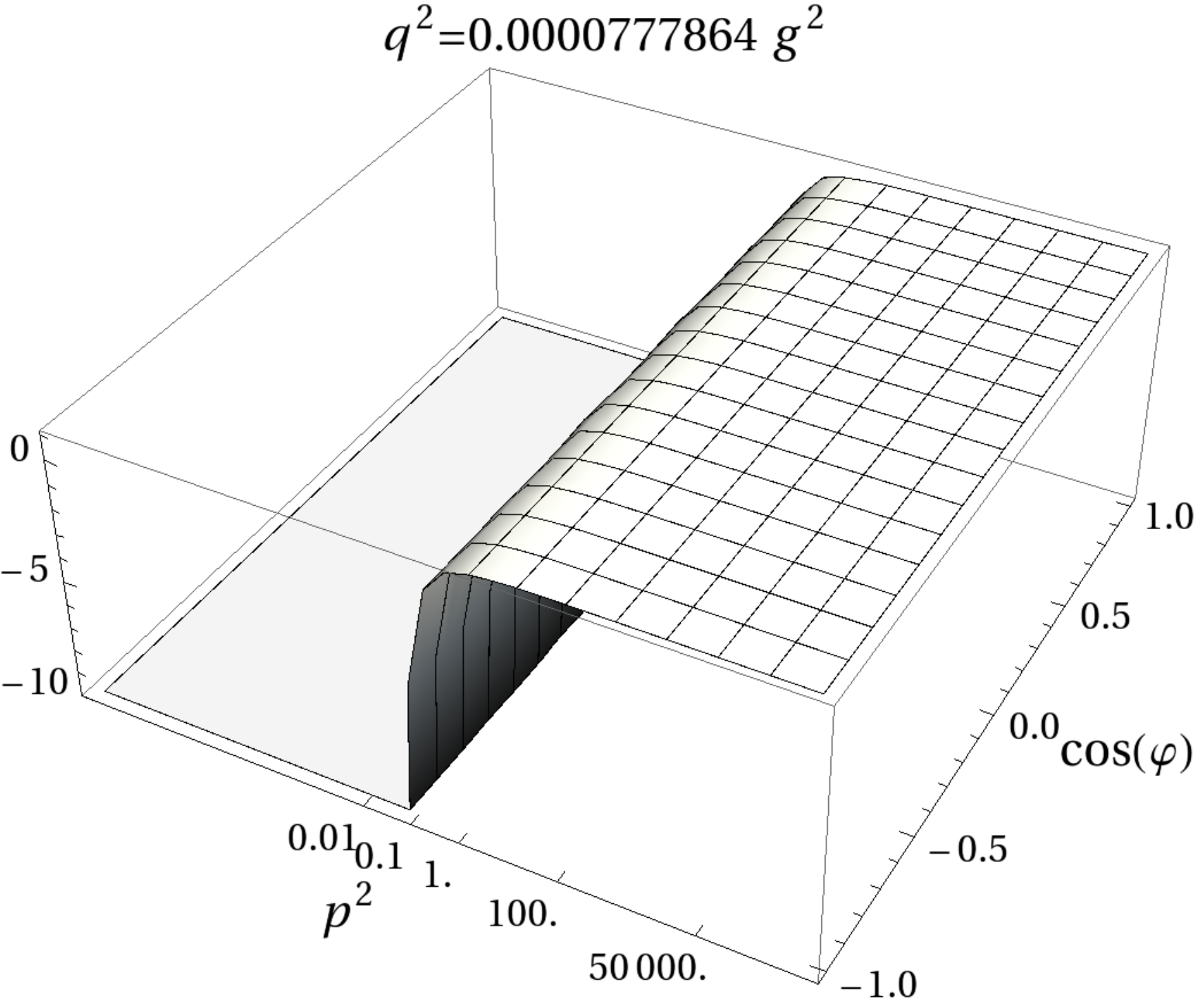}
 \includegraphics[width=\textwidth]{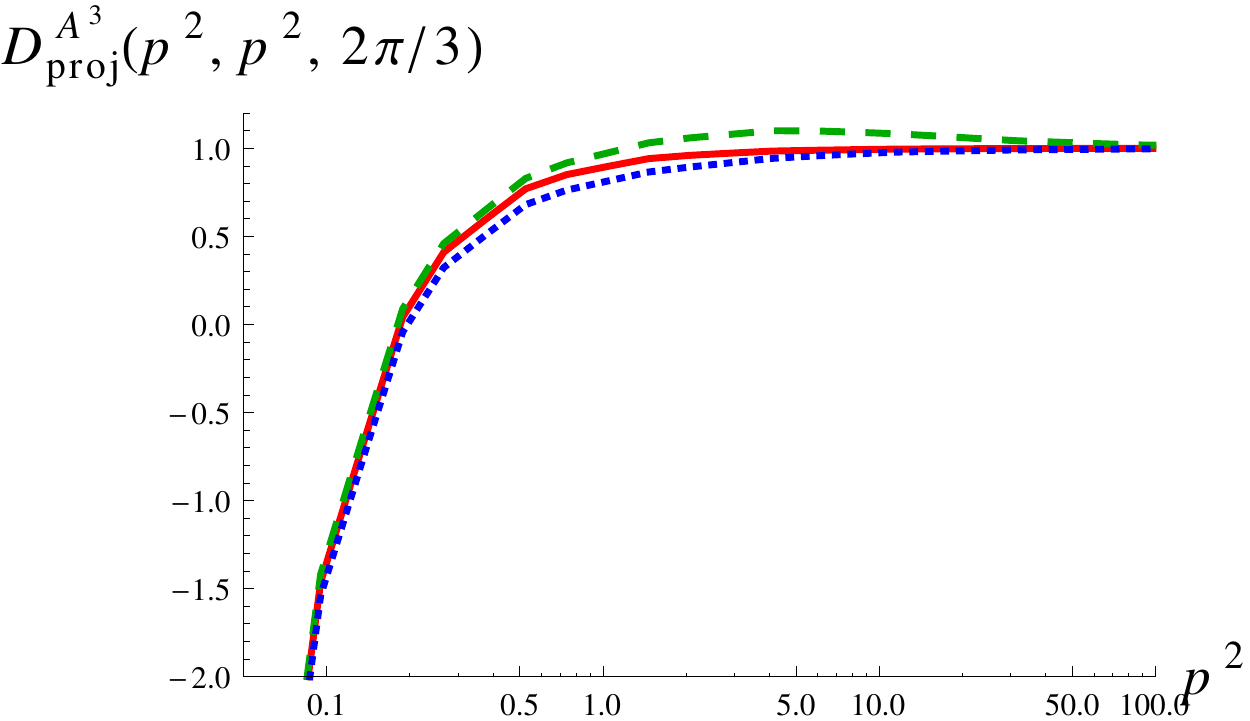}
 \end{minipage}
  \caption{\label{fig:3g-results}Dressing of the three-gluon vertex, see \eref{eq:tg_proj}. Only the beginning of the IR divergence is shown by cutting all data below $-10$. \textit{Top left:} Angle fixed at $\arccos(-0.41625)$. \textit{Top right:} One momentum is fixed at $\sqrt{0.00007786}\,g$.
  \textit{Bottom right:} Symmetric configuration. The solid/red line is with the ghost triangle only, the dashed/green one with both triangles and the dotted/blue line with all five diagrams.}
 \end{center}
\end{figure}

\begin{figure}[tb]
 \begin{center}
 \begin{minipage}[t]{0.48\textwidth}
 \includegraphics[width=\textwidth]{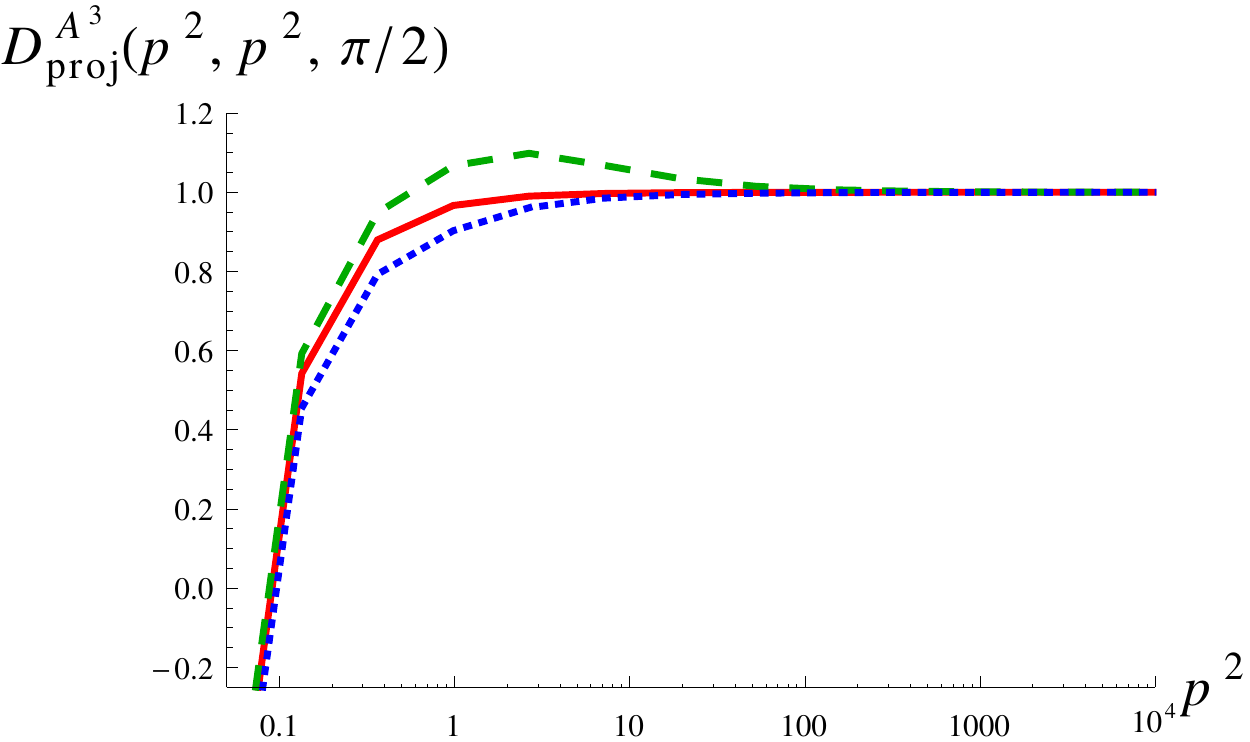}
 \end{minipage}
 \hfill
 \begin{minipage}[t]{0.48\textwidth}
 \includegraphics[width=\textwidth]{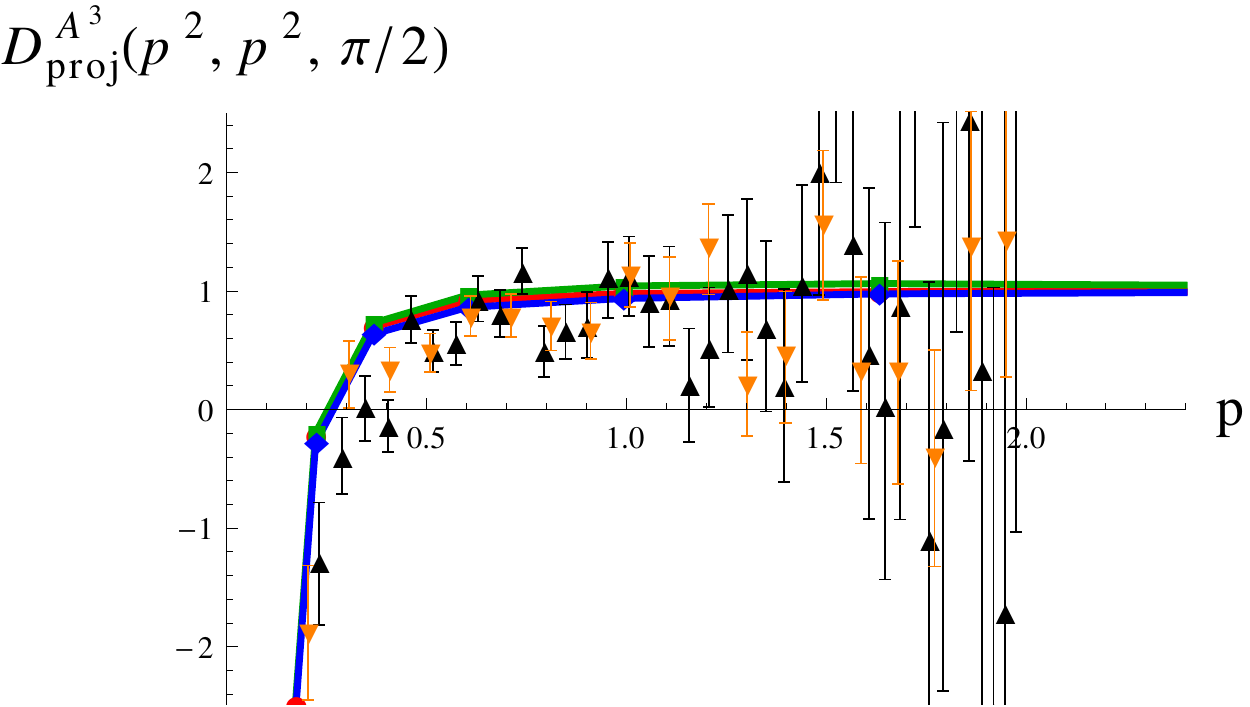}
 \end{minipage}
\caption{\label{fig:3g-results_ortho}Three-gluon vertex dressing for two orthogonal momenta. \textit{Left:} The red/solid curve results from using the ghost triangle alone, the green/dashed curve from ghost and gluon triangles and the blue/dotted curve from all five diagrams. The data is cut at the bottom to highlight the mid-momentum behavior. \textit{Right:} Comparison with lattice data. The red/solid, green/dashed and blue/dotted curves are from the DSE calculations as on the left-hand side, but with $SU(2)$, the dots are from lattice calculations \cite{Maas:2007uv}. Black up-triangles are for $\beta=10$/$L=21fm^{-1}$ and orange down-triangles for $\beta=22.5$/$L=12fm^{-1}$. The data is cut at the bottom and for the lattice data also in the UV, since the fluctuations are very large there.}
 \end{center}
\end{figure}

In lattice calculations the calculated quantity is the contraction of the transversely projected and amputated three-gluon vertex with the bare three-gluon vertex \cite{Cucchieri:2006tf,Maas:2007uv} normalized to the tree-level:
\begin{align}\label{eq:tg_proj}
 D^{A^3}_{proj}(p^2,q^2, \varphi):=\frac{\Gamma_{\mu\nu\rho}^{A^3,abc,(0)}(p,q,r) D_{gl,\mu\mu'}(p)D_{gl,\nu\nu'}(q)D_{gl,\rho\rho'}(r) \Gamma_{\mu'\nu'\rho'}^{A^3,abc}(p,q,r)}{\Gamma_{\mu\nu\rho}^{A^3,abc,(0)}(p,q,r) D_{gl,\mu\mu'}(p)D_{gl,\nu\nu'}(q)D_{gl,\rho\rho'}(r) \Gamma_{\mu'\nu'\rho'}^{A^3,abc,(0)}(p,q,r)}.
\end{align}
This quantity was calculated using a standard fixed-point iteration. The results for some selected momentum configurations are shown in \fref{fig:3g-results}. The qualitative features expected from lattice calculations, especially the zero crossing, are all there. At the symmetric point we compare the results when taking into account only the ghost triangle, both triangles or all five diagrams of the truncation described above. For the so-called orthogonal momentum configuration a comparison between continuum and lattice results is shown in \fref{fig:3g-results_ortho}, where again also the influence of the gluon triangle and the swordfish diagrams is shown. As expected the gluon triangle only gives a contribution in the mid-momentum regime and creates a small bump there which is absent when only the ghost triangle is taken into account. Interestingly, the swordfish diagrams reverse that effect and the bump goes away again. The zero crossing is also almost unaffected by the inclusion of further diagrams. Note that the small influence of gluons in the three-gluon vertex DSE is a non-trivial result insofar as for the ghost-triangle-only truncation the three-gluon vertex does not appear on the right-hand side and the iteration consists of only one step. Only the inclusion of the gluonic diagrams leads to the feedback of the three-gluon vertex onto itself. As expected, the details of the mid-momentum regime depend on the employed truncation, whereas the IR and UV regime are unaffected.

\bibliographystyle{utphys_mod}
\bibliography{literature_YM2d}

\end{document}